\documentclass[pra,twocolumn,longbibliography,amsmath,superscriptaddress,amssymb]{revtex4-1}

\usepackage{graphicx}
\usepackage{dcolumn}
\usepackage{bm}
\usepackage{amssymb}
\usepackage{amsmath}
\usepackage{color}
\usepackage[dvipsnames]{xcolor}
\usepackage{placeins}
\usepackage{epstopdf}

\usepackage[normalem]{ulem}

\usepackage[linkcolor=blue,urlcolor=blue,citecolor=blue,colorlinks=true]{hyperref}

\newcommand{\spr}{\shortparallel}
\newcommand{\rtm}[1]{\mathrm{#1}}

\newcommand{\Jexv}{\bm{\mathcal{J}}}
\newcommand{\Jex}{\mathcal{J}}

\begin{document}

\title{Variety of scenarios of the magnetic exchange response in topological insulators}

\author{I. A. Nechaev}
\affiliation{Donostia International Physics Center (DIPC), Paseo Manuel de Lardizabal 4, 20018 Donostia/San Sebasti\'{a}n,  Basque Country, Spain}
\affiliation{Department of Electricity and Electronics, FCT-ZTF, UPV-EHU, 48080 Bilbao, Spain}

\author{E. E. Krasovskii}
\affiliation{Donostia International Physics Center (DIPC), Paseo Manuel de Lardizabal 4, 20018 Donostia/San Sebasti\'{a}n,  Basque Country, Spain}
\affiliation{Departamento de Pol\'imeros y Materiales Avanzados: F\'isica, Qu\'imica y Tecnolog\'ia, Universidad del Pais Vasco/Euskal Herriko Unibertsitatea, 20080 Donostia/San Sebasti\'{a}n, Basque Country, Spain}
\affiliation{IKERBASQUE, Basque Foundation for Science, 48013 Bilbao, Basque Country, Spain}

\date{\today}

\begin{abstract}
We present an \textit{ab initio} relativistic $\mathbf{k}\cdot\mathbf{p}$  theory of the effect of magnetic exchange field on the band structure in the gap region of bulk crystals and thin films of three-dimensional layered topological insulators. For the field perpendicular to the layers (along $z$), we reveal novel unconventional scenarios of the response of the band-gap edges to the magnetization. The modification of the valence and conduction states is considered in terms of their $\Gamma$-point spin $s^z$ and total angular momentum $J^z$ on the atomic sites where the states are localized. The actual scenario depends on whether $s^z$ and $J^z$ have the same or opposite sign. In particular, the opposite sign for the valence state and the same sign for the conduction state give rise to an unconventional response in Bi$_2$Te$_3$---both in the bulk crystal and in ultra-thin films, which fundamentally distinguishes this topological insulator from Bi$_2$Se$_3$, where both states have the same sign. To gain a deeper insight into different scenarios in insulators with both inverted and non-inverted zero-field band structure, a minimal four-band third-order $\mathbf{k}\cdot\mathbf{p}$ model is constructed from first principles. Within this model, we analyze the field-induced band structure of the insulators and identify Weyl nodes that appear in a magnetic phase and behave differently depending on the scenario. We characterize the topology of the modified band structure by the Chern number $\mathcal{C}$ and find the unconventional response to be accompanied by a large Chern number $\mathcal{C}=\pm3$.
\end{abstract}

\maketitle

\section{\label{Intro}Introduction}

In insulating materials with spin-orbit interaction (SOI), a combination of magnetic exchange interaction (MEI) with topologically non-trivial band structure brings about fascinating phenomena intensively studied over the last decade. A vivid example is the quantum anomalous Hall (QAH) effect manifesting itself as a quantized Hall conductivity at zero external magnetic field~\cite{Weng_AP_2015, Liu_ARCMP_2016, Tokura_NRP_2019}.  Similar to the related quantum spin Hall effect (QSH)~\cite{Kane_PRL_2005_1, Kane_PRL_2005_2, Bernevig_Science_2006} the QAH effect was predicted theoretically~\cite{Haldane_PRL_1988, Onoda_PRL_2003}, and it was shown to be realized in HgTe/CdTe quantum wells (QWs) doped with Mn~\cite{Liu_PRL_2008} and in the films of the three-dimensional (3D) topological insulator (TI) Bi$_2$Se$_3$ doped with magnetic transition-metal impurities~\cite{Yu_Science_2010}.

\begin{figure}[tbp]
\centering
\includegraphics[width=\columnwidth]{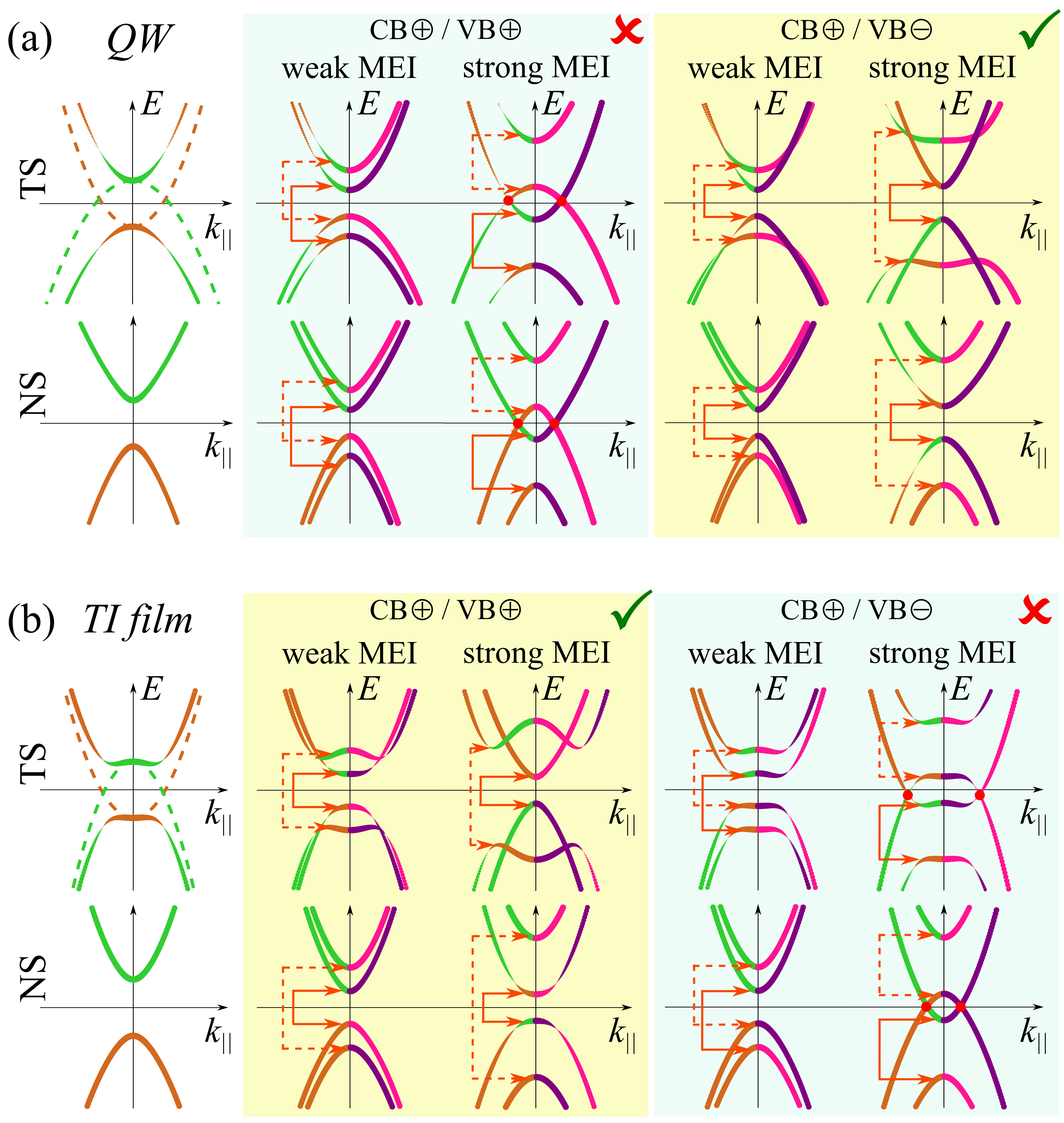}
\caption{Valence and conduction bands of quantum wells (a) and 3D TI films (b) around $\bar{\Gamma}$ upon the application of the $z$-directed exchange field within the four-band second-order $\mathbf{k}\cdot\mathbf{p}$ modeling of Refs.~\cite{Liu_PRL_2008} and \cite{Yu_Science_2010}, respectively. Encircled ``$+$'' and ``$-$'' mark the sign of the VB and CB splitting. Green and brown fat bands highlight the band inversion (in the leftmost column dashed lines illustrate the bands without SOI). Pink and purple fat bands show the model states with positive and negative $z$-\textit{pseudospin}, respectively. Orange arrows link the subbands of the TS or NS pairs (see text). The response of QWs and TI films that results in a QAH state is shown on the pale yellow background. The plots on the pale blue background illustrate the response that does not meet the conditions for the QAH effect, bringing the QWs and films to a gapless state due to the crossing (red points) of the subbands of the different pairs.}
\label{fig1}
\end{figure}

According to the theory of Refs.~\cite{Liu_PRL_2008,Yu_Science_2010}, for an insulating material to be brought to a QAH state by the MEI it must be close to the quantum phase transition, be it originally in a topologically non-trivial state (TS) or in a trivial state (NS). Without MEI, owing to the time-reversal symmetry (TRS), both NS and TS have Kramers-degenerate band structure, with the band-gap edges---the highest valence band (VB) and the lowest conduction band (CB)---being inverted in TS. In the minimal \textit{second-order} $\mathbf{k}\cdot\mathbf{p}$ models of Refs.~\cite{Liu_PRL_2008, Yu_Science_2010}, these edges are grouped into two pairs so that in the TS the gaps of both pairs are inverted, and each of them gives the quantum of the Hall conductivity with opposite sign $\pm e^2/h$ (the resulting two quanta add up to zero). Then, the QAH effect is realized if a finite MEI along $z$-axis, while breaking TRS, keeps the material insulating and is strong enough to either lift the inversion of one of the TS pairs or to invert one of the NS pairs~\cite{Yu_Science_2010, Liu_ARCMP_2016}, Fig.~\ref{fig1}. Then in the magnetic phase there is only one pair with the inverted gap, which leads to the quantized anomalous Hall conductivity.

The minimal $\mathbf{k}\cdot\mathbf{p}$ models reduce the effect of the $z$-directed magnetic exchange field to a Zeeman-type splitting of the initially degenerate CB and VB into subbands with \textit{up} and \textit{down} \textit{pseudospin} (the $z$-projection of the total angular momentum) by introducing a phenomenological exchange term with a separate parameter for each band-gap edge. As a consequence, the behavior of a TS or NS pair depends upon two exchange parameters  because each pair includes one VB and one CB subband. These parameters account for the strength and sign of the splitting: A positive value causes the \textit{pseudospin-up} subband of the VB or CB to increase (\textit{pseudospin-down} to decrease) its energy, which is referred to as a positive splitting. A negative splitting implies a decrease of the \textit{pseudospin-up} (increase of the \textit{pseudospin-down}) subband energy, see Fig.~\ref{fig1}.

To meet the conditions for the QAH effect, the MEI must differently affect the gaps of the pairs. For the HgTe/CdTe QWs, where both subbands of each pair have the same \textit{pseudospin}, Fig.~\ref{fig1}(a), this means that the magnetic splitting is of opposite sign for VB and CB~\cite{Liu_PRL_2008}, while in the magnetically doped TI films, where each pair comprises subbands with opposite \textit{pseudospins}, the splitting is of the same sign~\cite{Yu_Science_2010}, Fig.~\ref{fig1}(b). The latter justifies using the traditional Zeeman-like term $\sigma_z M_z$ for the magnetic TIs, where $M_z$ is the $z$-directed magnetic exchange field and $\sigma_z$ is the Pauli matrix acting in the \textit{pseudospin} space and commonly associated with the \textit{real spin}~\cite{Lu_PRL_2013, Wang_PRL_2015, Zhang_PRL_2020}. Apart from assuming the same signs of the VB and CB splitting, this simplification also implies the same magnitude of splitting for these bands.

However, in Ref.~\cite{Zhang_Science_2013}, to reach an agreement between the theory based on the four-band $\mathbf{k}\cdot\mathbf{p}$ model of Ref.~\cite{Zhang_NATPHYS_2009} and transport measurements in Cr-doped Bi$_2$(Se$_x$Te$_{1-x}$)$_3$ films, the VB and CB splittings of the bulk Bi$_{1.75}$Cr$_{0.25}$(Se$_x$Te$_{1-x}$)$_3$  were heuristically chosen to be of opposite sign. [This implies that in the $k_z=0$ plane the bulk VB and CB respond to the MEI similarly to the TS pairs shown in the pale blue panel of Fig.~\ref{fig1}(b).] The splitting magnitudes were estimated to be rather small compared to the band gap, thereby excluding a quantum phase transition driven by the exchange field. However, the physics behind the choice of opposite signs for the 3D TIs and relevant scenarios of the MEI effect on their topological and spin-related properties have not been addressed so far. In the spirit of the pioneering studies ~\cite{Kane_PRL_2005_1, Kane_PRL_2005_2, Bernevig_Science_2006, Liu_PRL_2008, Zhang_NATPHYS_2009, Yu_Science_2010}, we will construct an effective model to address all possible splitting scenarios, with the parameters related to the \textit{real spin} being unambiguously derived from the {\it ab initio} band structure.

In this paper, we study the effect of the MEI applied along $z$-axis on the bulk crystals of Bi$_2$Se$_3$, Bi$_2$Te$_2$Se, Bi$_2$Te$_3$, and Sb$_2$Te$_3$ as well as on their thin films [up to six quintuple layers (QLs)] with effective \textit{third-order} $\mathbf{k}\cdot\mathbf{p}$ Hamiltonians. The Hamiltonians are obtained using the \textit{ab initio} relativistic $\mathbf{k}\cdot\mathbf{p}$ theory, which has been successfully applied to both magnetic and non-magnetic materials with SOI~\cite{Nechaev_PRBR_2016, Nechaev_SciRep_2017, Nechaev_PRB_2018, Susanne2019, Usachov_PRL_2020} and demonstrated to reliably describe subtle spin-related effects, such as the sign reversal of the spin polarization within a non-degenerate band~\cite{Nechaev_PRB_2019} and non-orthogonality between momentum and spin~\cite{Nechaev_PRB_2020}. The models are derived immediately from the eigenfunctions of the original \textit{ab initio} Hamiltonian and, therefore, do not involve adjustable parameters. This allows us to establish the actual relation between the \textit{pseudospin} $J^z$  and \textit{real spin} $s^z$ of the subbands and, consequently, to simulate the MEI effect of different strength by introducing an exchange term that influences the subbands according to their \textit{real spins}. We show that in the bulk crystals of Bi$_2$Se$_3$ and Bi$_2$Te$_2$Se the co-directional $z$-projections of \textit{real spin} and \textit{pseudospin} imply a conventional response to the MEI (splitting of VB and CB have the same sign), while an unconventional response (opposite sign for VB and CB) takes place in Bi$_2$Te$_3$ and Sb$_2$Te$_3$ due to the opposite sign of $J^z$ and $s^z$ for the VB.

To illustrate possible scenarios of the MEI effect, we construct a minimal $\mathbf{k}\cdot\mathbf{p}$ model and apply it to hypothetical topologically trivial and non-trivial bulk insulators. The model employs a \textit{microscopically} derived four-band third-order Hamiltonian, which is an improvement over the model third-order Hamiltonian of Ref.~\cite{Liu_PRB_2010} and therefore allows us to reveal new topological features of 3D TI-like magnetic materials. We calculate the Chern number $\mathcal{C}(k_z)$ as a function of the Bloch vector $k_z$ along the exchange-field direction and demonstrate that in the topologically trivial insulators there emerges a phase with a nonzero $\mathcal{C}(k_z)$ when a sufficiently strong MEI induces the inversion of the band-gap edges. Here, the conventional response to the exchange field leads to a nontrivial topological band structure, which, in addition to a pair of Weyl nodes along $k_z$-axis, may also have Weyl nodes in the mirror plane away from high symmetry lines in the Brillouin zone (BZ). Due to the symmetry, these mirror-plane nodes lead to jumps in the Chern number by $\pm3$. In the unconventional response, there are no Weyl nodes in the  $k_z$-axis, and a nonzero $\mathcal{C}(k_z)$ larger than in the conventional case is entirely due to the mirror-plane nodes. The non-trivial insulators differ from the trivial ones in that the mirror-plane Weyl nodes that cause a nonzero $\mathcal{C}(k_z)$ appear already at a weak MEI well before the band inversion is lifted.

For the films of the real TIs, we find the unconventional response to the exchange field only in ultrathin films of Bi$_2$Te$_3$. In particular, we show that the two-QL Bi$_2$Te$_3$ film in a strong exchange field transfers from a QSH state to a large-Chern-number QAH state ($\mathcal{C}=-3$). This corresponds to the strong MEI case presented in the pale blue panel of Fig.~\ref{fig1}(b), when the field does not lift the inversion of any of the TS pairs but induces a crossing of two subbands of different TS pairs (red points in the figure). However, a more accurate treatment of the interaction between these subbands results in an avoided-crossing gap. Thicker Bi$_2$Te$_3$ films and films of the other TIs demonstrate the conventional response leading to a phase transition to a QAH state with $\mathcal{C}=1$ in accord with the theory of Ref.~\cite{Yu_Science_2010}.

\section{\label{Comp_Det}Computational details}

Similar to Ref.~\cite{Nechaev_PRBR_2016}, the \textit{ab initio} band structure is obtained with the ELAPW method~\cite{Krasovskii_PRB_1997} using the full potential scheme of Ref.~\cite{Krasovskii_PRB_1999} within the local density approximation (LDA). The spin-orbit interaction is treated by a second variation method~\cite{Koelling_1977}. The experimental crystal lattice parameters of the 3D topological insulators (TIs) were taken from Ref.~\cite{Wyckoff_RWG}. In the case of Bi$_2$Te$_2$Se, the experimental atomic positions of Ref.~\cite{Wyckoff_RWG} were used, while for Bi$_2$Se$_3$, Bi$_2$Te$_3$, and Sb$_2$Te$_3$ we took the LDA relaxed atomic positions of Refs.~\cite{Nechaev_PRB_2013_BISE, Nechaev_PRB_2013_BITE, Nechaev_PRB_2015_SBTE}. The bulk crystals have space group $R\bar{3}m$ (no.~166), and thin films are bulk-truncated centrosymmetric repeated slabs of space group $P\bar{3}m1$ (no.~164). The thickness of the films varies from one to six QLs.

The \textit{ab initio} relativistic $\mathbf{k}\cdot\mathbf{p}$ perturbation expansion is carried out at the center of the BZ up to the third order in $\mathbf{k}$ by applying the L\"{o}wdin partitioning~\cite{Leowdin_JCP_1951, Schrieffer_PR_1966, Winkler_KP} to the original Hilbert space of the $\Gamma$-projected all-electron Hamiltonian $\mathcal{H}_{{\mathbf k}}$ (see Ref.~\cite{Nechaev_PRB_2020}). The size of the ${\mathbf k}\cdot{\mathbf p}$ Hamiltonian $H_{\rtm{\mathbf{kp}}}$ is determined by the number of basis functions---the Kramers pairs $\Psi_{n\mu}$ of a set of doubly degenerate levels $\epsilon_n$. Thus, the basis of our ${\mathbf k}\cdot{\mathbf p}$ models is a unitary transformed fragment of the original all-electron spectrum at $\Gamma$ with the subscript $\mu=\uparrow$ or $\downarrow$ indicating the sign of the \textit{on-site} expectation value of the $z$ component $\widehat{J}_z$ of the total angular momentum $\widehat{\mathbf{J}} = \widehat{\mathbf{L}} + \widehat{\mathbf{S}}$. This expectation value is evaluated at the (symmetry equivalent) atomic sites, which have the largest weight in the $n$-th level (see Appendix~\ref{A_onsite} for details). The subscript $\mu$ indicates a member of the Kramers pair, and it is omitted for brevity when a discrimination between the members of the pair is unimportant. For a \textit{minimal} $\mathbf{k}\cdot\mathbf{p}$ model we use a basis of only two Kramers-degenerate pairs, namely the \textit{highest} valence and \textit{lowest} conduction states at $\Gamma$ (hereafter referred to simply as the valence and conduction states).

\begin{figure}[tbp]
\centering
\includegraphics[width=\columnwidth]{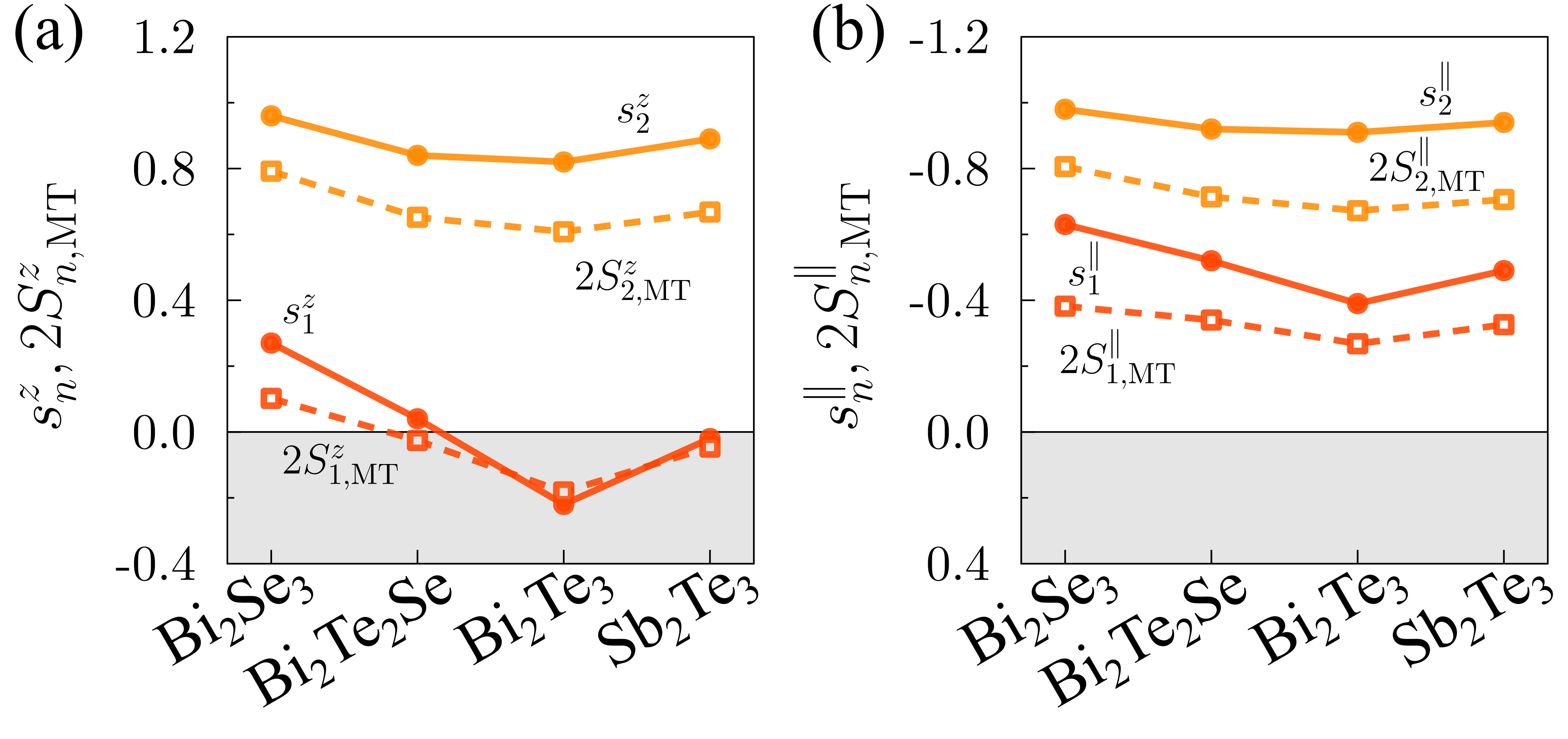}
\caption{(a) Expectation values $s_n^z = \langle\Psi_{n\uparrow} | \sigma_z | \Psi_{n\uparrow}\rangle$ in the basis state $n=1$ and 2. The sum of the \textit{on-site} spin $z$-projection expectation values $S_{n,\rtm{MT}}^z = \sum_i \frac{1}{2}\langle\Psi_{n\uparrow} | \sigma_z | \Psi_{n\uparrow}\rangle_i$ over all atomic sites in the unit cell is also shown. (b) Same as in graph (a), but for $s_n^{\spr} = \langle\Psi_{n\uparrow} | \sigma_x | \Psi_{n\downarrow}\rangle$ and $S_{n,\rtm{MT}}^{\spr} = \sum_i \frac{1}{2}\langle\Psi_{n\uparrow} | \sigma_x | \Psi_{n\downarrow}\rangle_i$.}
\label{fig2}
\end{figure}

Following Refs.~\cite{Nechaev_PRB_2018} and \cite{Usachov_PRL_2020}, we model hypothetical magnetic phases by the  $\mathbf{k}\cdot\mathbf{p}$ Hamiltonian $H_{\rtm{\mathbf{kp}}} + H_{\rtm{EX}}$ with the matrix elements of the additional exchange-filed contribution $\left[H_{\rtm{EX}}\right]^{n\mu}_{l\nu} = -\Jexv_{nl}\cdot \left[ \rtm{\mathbf{S}}_{\rtm{\mathbf{kp}}} \right]^{n\mu}_{l\nu}$ expressed in terms of the spin matrix elements $\left[\rtm{\mathbf{S}}_{\rtm{\mathbf{kp}}} \right]^{n\mu}_{l\nu} = \langle\Psi_{n\mu}| \bm{\sigma} |\Psi_{l\nu}\rangle$, where $\bm{\sigma}=(\sigma_x,\sigma_y,\sigma_z)$ is the vector of the Pauli matrices. Here, the exchange-interaction parameters are given by the vector $\Jexv_{nl}$ ($\Jexv_{ln}=\Jexv_{nl}$) collinear to the magnetization $\rtm{\mathbf{M}}$ due to ferromagnetic doping. In this study, we consider the out-of-plane (along $z$) orientation of the magnetization,  $\rtm{\mathbf{M}} \uparrow \uparrow \widehat{\mathbf{z}}$. For simplicity, we assume the exchange-interaction parameters to be nonzero only for the valence and conduction states and to be the same for the two states, i.e., $\Jexv_{nl} = \Jex\widehat{\mathbf{z}}$ for $n,l\in\{1,2\}$ and $\Jexv_{nl}=0$ for $n,l\notin\{1,2\}$. (These states are numbered in order of increasing energy: the valence state $|\Psi_{1}\rangle$ and the conduction state $|\Psi_{2}\rangle$.) As a consequence, the nonzero part of $H_{\rtm{EX}}$ is given by a $4\times4$ matrix, the corresponding $4\times4$ block of the spin matrix being
\begin{equation}\label{kpSpinBulk}
\rtm{\mathbf{S}}^{4\times4}_{\rtm{\mathbf{kp}}}=\left(
\begin{array}{cc}
\rtm{\mathbf{S}}_1   & 0       \\
0                                       & \rtm{\mathbf{S}}_2
\end{array}
\right).
\end{equation}
Here $\rtm{\mathbf{S}}_n=(s^{\spr}_n\bm{\sigma}_{\spr},s^{z}_n\sigma_z)$ with $\bm{\sigma}_{\spr} = (\sigma_x, \sigma_y) $ and $n=1$ or 2. Thus, the effect of the $z$-directed exchange field on the TI band gap is described by the matrix $H^{4\times4}_{\rtm{EX}} =-\Jex\widehat{\mathbf{z}} \cdot \rtm{\mathbf{S}}^{4\times4}_{\rtm{\mathbf{kp}}}$ proportional to the exchange-interaction parameter $\Jex$ and to the spin expectation values $s^{z}_{1,2}$.

\section{\label{onsite_sec}Language of Atomic Orbitals}

\begin{figure}[tbp]
\centering
\includegraphics[width=\columnwidth]{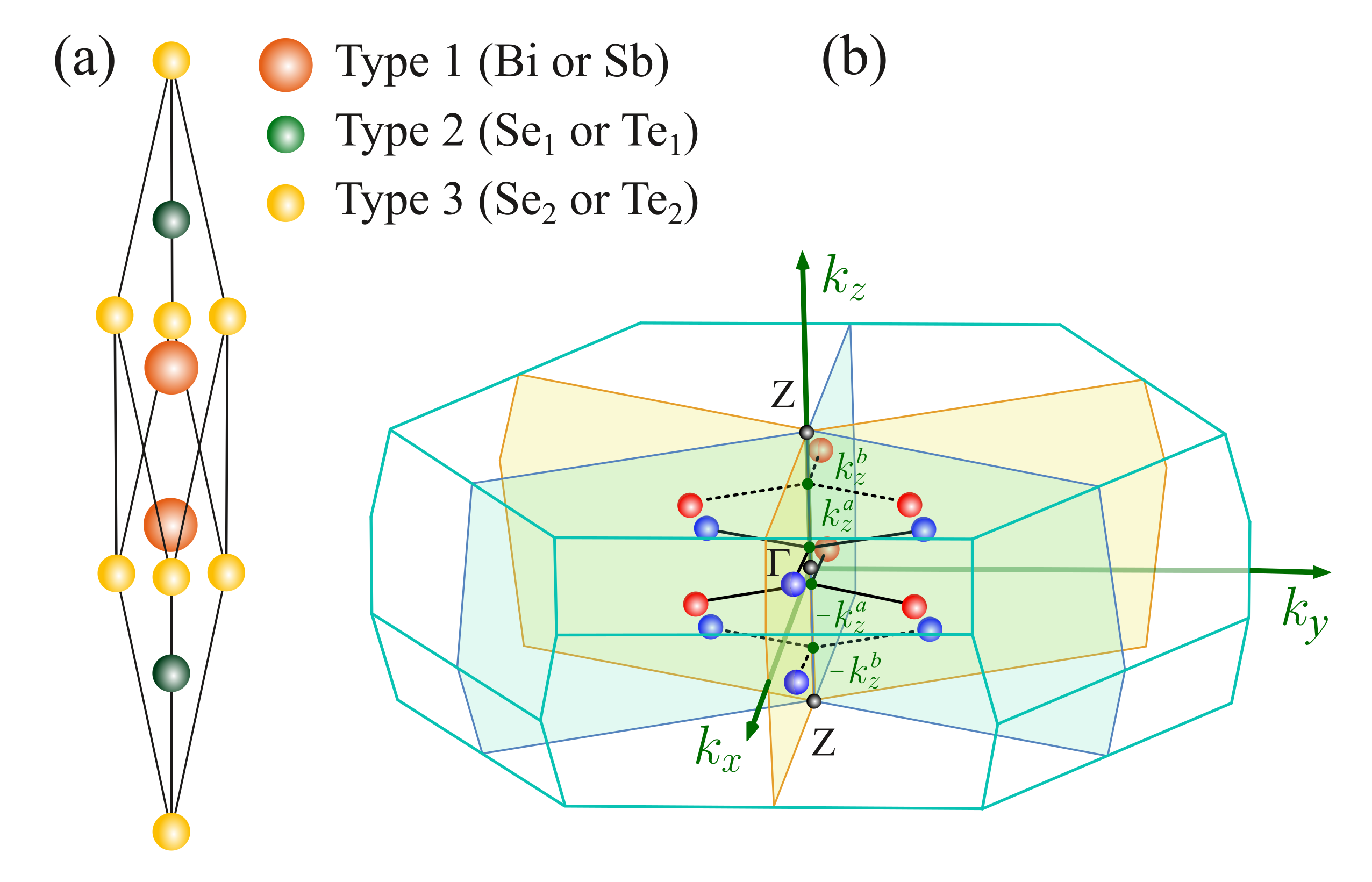}
\caption{Rhombohedral unit cell (a) and BZ (b) of the bulk TIs. In graph (b), the Weyl nodes appearing in the mirror planes (pale orange and pale blue areas) due to the $z$-directed exchange field are shown by red (positive chiral charge) and blue (negative chiral charge) balls. In constant-$k_z$ planes, the three nodes and their partners are connected by the rays of the same style (solid or dashed).}
\label{fig3}
\end{figure}

In the effective $\mathbf{k}\cdot\mathbf{p}$ approach to magnetism in TIs, the common strategy is to consider the basis wave function as if they were eigenfunctions of the spin operator along $\mathbf z$~\cite{Silvestrov_PRB_2012, Brey_PRB_2014, Zhang_PRL_2020}. In Eq.~(\ref{kpSpinBulk}), this corresponds to $s^{z}_{1} = s^{z}_{2} = s^{\spr}_{1} = s^{\spr}_{2} = 1$. This is, however, not the case in the {\it ab initio} calculations, Fig.~\ref{fig2}: while $s_2^z$ (CB) is almost unity for all the bulk TIs studied, the parameter $s_1^z$ (VB) is significantly smaller than unity, and it has different magnitude and even different sign in different TIs [see Fig.~\ref{fig2}(a)].

\begin{figure*}[tbp]
\centering
\includegraphics[width=\textwidth]{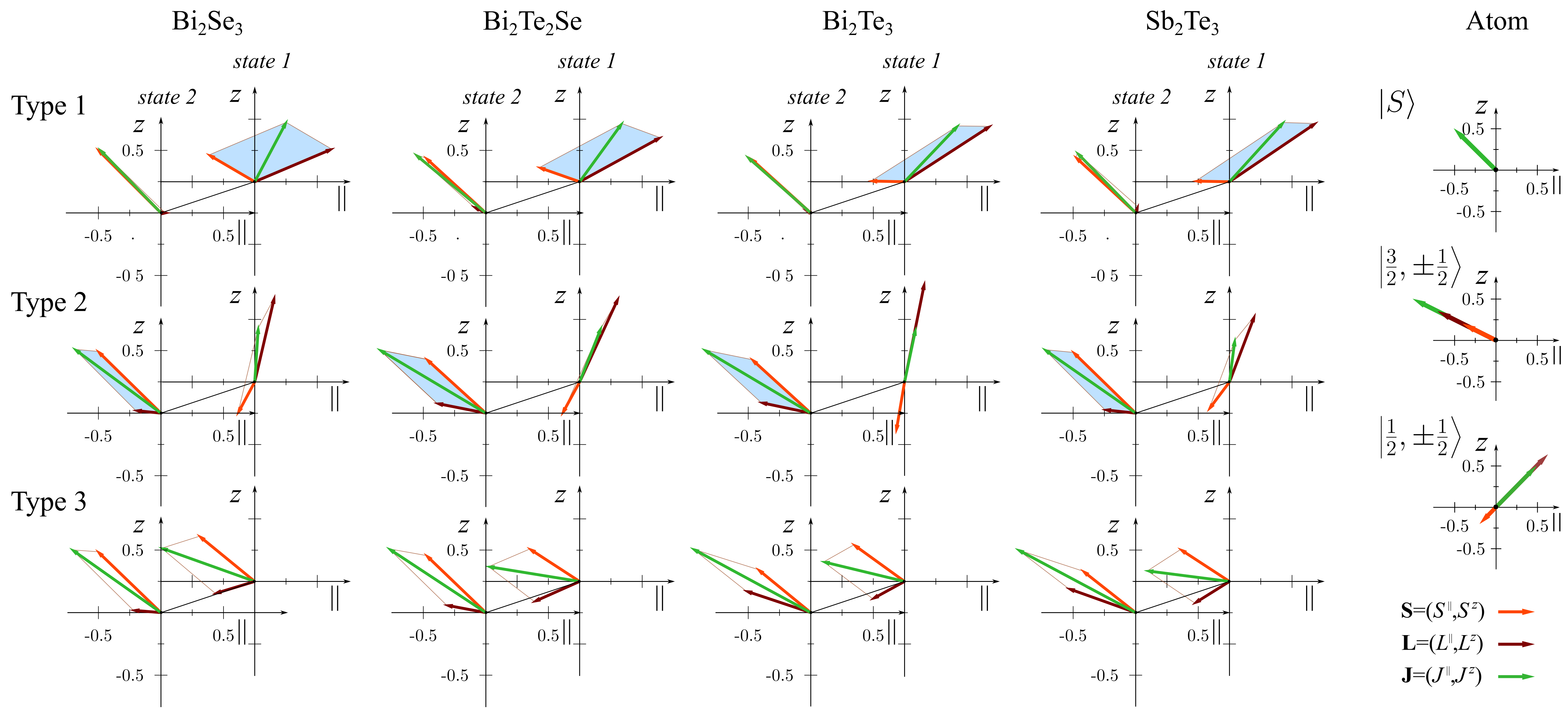}
\caption{On-site expectation values of the orbital angular momentum $\widehat{\mathbf{L}}$, spin $\widehat{\mathbf{S}}$, and total angular momentum $\widehat{\mathbf{J}}$ in the valence ($n=1$) and the conduction ($n=2$) state for an atom of each type. The values are presented by the vectors $\mathbf{L}_{n,i}$, $\mathbf{S}_{n,i}$, and $\mathbf{J}_{n,i}$ on the two-dimensional ($\spr,\,z$) plane (see Appendix~\ref{A_onsite}). For each state, the pale blue parallelogram is formed by the vectors $\widehat{\mathbf{L}}$ and $\widehat{\mathbf{S}}$ evaluated in a sphere of the type that has the largest weight in this state. The rightmost column shows the expectation values in the atomic states of Eq.~(\ref{fig4}).}
\label{LSJ_on_site}
\end{figure*}

To see the actual spin distribution behind $s^{z}_1$ and $s^{z}_2$, let us consider the \textit{on-site} spin $z$ contributions and their relation to the \textit{on-site} expectation values of the orbital angular momentum $\widehat{\mathbf{L}}$ and total angular momentum $\widehat{\mathbf{J}}$. The \textit{on-site} value is defined as the expectation value of an operator in an eigenstate of $\mathcal{H}_{{\mathbf k}}$ at $\Gamma$ calculated as an integral over the muffin-tin (MT) sphere of the atomic site, see Appendix~\ref{A_onsite}. In the bulk crystals, there are three types of the symmetry equivalent sites, $\tau=1$, 2, and 3 [see Fig.~\ref{fig3}(a)]. The types 1 and 2 comprise two sites each: $\tau=1$ is Bi or Sb in the planes next to the middle plane of the QL and $\tau=2$ is Se$_1$ or Te$_1$ in the QL outer planes. The type 3 has one site, Se$_2$ or Te$_2$ in the middle plane of the QL.

In Fig.~\ref{fig2}(a), we show the contribution to $s_n^z$ from all the MT spheres. For example, for Bi$_2$Se$_3$ the MT-contribution is $S_{n,\rtm{MT}}^z = 2S_{n,\rtm{Bi}}^z + 2S_{n,\rtm{Se_1}}^z + S_{n,\rtm{Se_2}}^z$ with $S_{n,i}^z$ listed in Table~\ref{tab:table1} of Appendix~\ref{A_onsite}. The appreciable difference between $s_n^z$ and the  MT-contribution to it is obviously due to the spin density in the interstitial region. As seen in Fig.~\ref{fig2}(a), $s_2^z$ is dominated by the MT-contribution; it is positive and varies slightly from one TI to another. By contrast, $s_1^z$ behaves differently in different materials: it is positive for Bi$_2$Se$_3$ and Bi$_2$Te$_2$Se but negative for Bi$_2$Te$_3$ and Sb$_2$Te$_3$.  The contributions of the MT-spheres and the interstitial region to $s_1^z$ have the same sign in Bi$_2$Se$_3$ and Bi$_2$Te$_3$, but for Bi$_2$Te$_2$Se and Sb$_2$Te$_3$ the interstitial and MT contributions have opposite signs, and the resulting total spin $s_1^z$ is close to zero.

It is instructive to compare $s^z$ and $s^{\spr}$ for different materials,  see Fig.~\ref{fig2}(b). For $|\Psi_{2}\rangle$ (CB), the relation between $s_2^{\spr}$ and $S_{2,\rtm{MT}}^{\spr}$ is similar to that of $s_2^{z}$ and $S_{2,\rtm{MT}}^{z}$ [Fig.~\ref{fig2}(a)], which means that also for the ${\spr}$-spin the interstitial and MT contributions are of the same sign. For $|\Psi_{1}\rangle$ (VB), the MT  and the interstitial contribution to $s_1^\spr$ are also of the same sign for all TIs, while this is not the case for $s_1^z$. Thus, the material's individuality manifests itself in $s^z$ but not in $s^{\spr}$.

In Fig.~\ref{LSJ_on_site}, we visualize the on-site expectation values of the operators $\widehat{\mathbf{L}}$, $\widehat{\mathbf{S}}$, and $\widehat{\mathbf{J}}$ by the vectors $\mathbf{L}_{n,i}$, $\mathbf{S}_{n,i}$, and $\mathbf{J}_{n,i}$ on the two-dimensional ($\spr,\,z$) plane, as explained in Appendix~\ref{A_onsite}. Keeping in mind that crystal breaks the rotation symmetry of a free atom, let us see if we still can use the atomic-orbital terminology $|jm_j\rangle$ for the Bloch states under study. Clearly, if the basis spinor wave functions were eigenfunctions of the operators $\widehat{\mathbf{J}}^2$ and $\widehat{J}_z$, the respective eigenvalues $j$ and $m_j$ would not depend on the volume of integration. Since the $p$ orbitals (mostly $p_z$) strongly dominate in the MT spheres that have the largest weight in the states $|\Psi_{1}\rangle$ and $|\Psi_{2}\rangle$, the deviation of the components of the vectors shown in Fig.~\ref{LSJ_on_site} (marked by pale blue parallelograms) from the values characteristic of the atomic states of $p$-character (the rightmost column in Fig.~\ref{LSJ_on_site}) is a measure of the departure from the atomic behavior.

As the reference atomic states, we consider the eigenfunctions of the total angular momentum $|jm_j\rangle$ widely used in $\mathbf{k}\cdot\mathbf{p}$ theory~\cite{Eppenga_PRB_1987, Foreman_PRB_1993, Winkler_KP, Dargys_2007, Abolfath_PRB_2011}:
\begin{eqnarray}\label{fig4}
\left|\frac{1}{2},\,\frac{1}{2}\right\rangle&=&\left(
\begin{array}{c}
s \\
0
\end{array}
\right),\,  \left|\frac{1}{2},-\frac{1}{2}\right\rangle=-\left(
\begin{array}{c}
0 \\
s
\end{array}
\right), \\
\left|\frac{1}{2},\,\frac{1}{2}\right\rangle &=& \frac{1}{\sqrt{3}}\left(
\begin{array}{c}
z \\
x+iy
\end{array}
\right),\, \left|\frac{1}{2},-\frac{1}{2}\right\rangle = \frac{1}{\sqrt{3}}\left(
\begin{array}{c}
x-iy\\
-z
\end{array}
\right),\nonumber\\
 \left|\frac{3}{2},\,\frac{1}{2}\right\rangle &=& \frac{1}{\sqrt{6}}\left(
\begin{array}{c}
-2z\\
x+iy
\end{array}
\right),\, \left|\frac{3}{2},-\frac{1}{2}\right\rangle = \frac{1}{\sqrt{6}}\left(
\begin{array}{c}
x-iy\\
2z
\end{array}
\right),\nonumber
\end{eqnarray}
where $s$, $x$, $y$, and $z$ are real $s$ and $p$ orbitals. The functions in Eq.~\ref{fig4} are for $m_j=\pm\frac{1}{2}$~\cite{Liu_PRB_2010}, with the phase factors being chosen such that the pairs are Kramers conjugate: $\left|j,-\frac{1}{2}\right\rangle = \hat{T}\left|j,\frac{1}{2}\right\rangle$, similar to our basis wave functions $|\Psi_{n\mu}\rangle$. Here $\hat{T} = Ki\sigma_y$ is the time reversal operator, and $K$ is the complex conjugation operator.

As seen in Fig.~\ref{LSJ_on_site}, all the four TIs show similar pattern caused by the similar orbital character of the states $|\Psi_{1}\rangle$ and $|\Psi_{2}\rangle$ in the respective spheres. For $|\Psi_{1}\rangle$, the type-1 vectors $\mathbf{L}_{1,i\in1}$ and $\mathbf{S}_{1,i\in1}$ make a large angle so that their $\spr$-projections have opposite signs and both $z$-projections are positive, Fig.~\ref{LSJ_on_site}. Due to the large orbital momentum $\mathbf{L}_{1,i\in1}$, the resulting vector $\mathbf{J}_{1,i\in1}$ gravitates towards $\mathbf{L}_{1,i\in1}$. Note that the $z$-projection of $\mathbf{J}_{1,i\in1}$ is nearly $\frac{1}{2}$, which is the $\widehat{J}_z$ expectation value in the $m_j=\pm\frac{1}{2}$ states. Also the value of $j$ found from the equation $j(j+1)=J^2_{1,i\in1}/Q_{1,i\in1}$ is about $0.7$, which is rather close to $j=\frac{1}{2}$. In contrast, for type~2, the angle between $\mathbf{L}_{1,i\in2}$ and $\mathbf{S}_{1,i\in2}$ is almost $\pi$, and their $z$-projections have opposite signs too. The total angular momentum $\mathbf{J}_{1,i\in2}$ deviates only slightly from $\mathbf{L}_{1,i\in2}$ as in the atomic states $\left|\frac{1}{2},\pm\frac{1}{2}\right\rangle$, see the rightmost column in Fig.~\ref{LSJ_on_site}. The difference from the atomic case is mainly due to the orbitals $p_{\pm}=p_x\pm ip_y$ dominating over $p_z$. Type~3 has a minor weight in the state, and we just note that both $\mathbf{L}_{1,i\in3}$ and $\mathbf{S}_{1,i\in3}$ have negative $\spr$-projections, while their $z$-projections are negative and positive, respectively.

In $|\Psi_{2}\rangle$, the orbital angular momentum $\mathbf{L}_{2,i\in1}$ is very small, so $\mathbf{J}_{2,i\in1}$ deviates only slightly from the respective spin $\mathbf{S}_{2,i\in1}$, which is practically the same for all TIs, with the $z$-projection of $\mathbf{J}_{2,i\in1}$ being very close to $\frac{1}{2}$. This situation is close to one of the atomic $|S\rangle$ states, see Fig.~\ref{LSJ_on_site}. For $\tau=2$, the spin $\mathbf{S}_{2,i\in2}$ behaves similarly, but $\mathbf{L}_{2,i\in2}$ is rather large while its $z$-projection is small. As a result, $\mathbf{J}_{2,i\in2}$ and $\mathbf{S}_{2,i\in2}$ are far from being co-directional. Although $J^z_{2,i\in2}$ is very close to $\frac{1}{2}$ and $j\approx1.4$ (from the ratio $J^2_{2,i\in2}/Q_{2,i\in2}$), the vectors $\mathbf{L}_{2,i\in2}$, $\mathbf{S}_{2,i\in2}$, and $\mathbf{J}_{2,i\in2}$ do not closely follow their atomic counterparts $|\frac{3}{2}, \pm \frac{1}{2} \rangle$ mainly due to the $p_z$ contribution being considerably larger than those of  $p_{\pm}$, Fig.~\ref{LSJ_on_site}. Finally, the vectors $\mathbf{L}_{2,i\in3}$, $\mathbf{S}_{2,i\in3}$, and $\mathbf{J}_{2,i\in3}$ of type~3 behave rather similar to those of type~2. The largest differences are observed in Sb$_2$Te$_3$.

Fig.~\ref{LSJ_on_site} clearly demonstrates that the Bloch states $|\Psi_{1}\rangle$ and $|\Psi_{2}\rangle$ near the nuclei are rather far from atomic-like: Unlike the pure atomic states, total angular momentum and spin are neither parallel nor antiparallel. At the same time, in the MT-spheres with the largest weight in $|\Psi_{1}\rangle$ and $|\Psi_{2}\rangle$, these Bloch states are characterized by the values of $J^z_{n,i}/Q_{n,i}$ that are rather close to $\frac{1}{2}$. Therefore, our choice of the Kramers pairs, with the members of the pair numbered by the subscript $\mu$ indicating the \textit{on-site} expectation value $J^z$ of the operator $\widehat{J}_z$, is in line with the traditional effective models of the $\mathbf{k}\cdot\mathbf{p}$ theory in the $|j,m_j\rangle$ representation. In practice, this choice turns out to preserve the unified form of the \textit{ab initio} derived $\mathbf{k}\cdot\mathbf{p}$ Hamiltonian for the whole class of TIs.

Turning back to the spin parameter $s^{z}$, we can conclude that in the state $|\Psi_{1}\rangle$ of Bi$_2$Te$_3$ the rather large \textit{negative} value of $s_1^{z}$ is due to the sizable negative $z$ spin on the Te-atoms of the QL outer planes and the nearly zero $z$ spin on the Bi atoms. In the state $|\Psi_{2}\rangle$ of Bi$_2$Te$_3$, the same Te-atoms are responsible for the {\em positive} $s_2^{z}$. This means that here the response of the conduction and valence state to the out-of-plane magnetization is almost exclusively determined by the Te-atoms of the QL outer planes or, very roughly, by their atomic-like $p$-states $|\frac{1}{2}, \pm \frac{1}{2} \rangle$ and $|\frac{3}{2}, \pm \frac{1}{2} \rangle$.

\section{\label{bulkEF_sec}External exchange field effect}

\begin{figure*}[tbp]
\centering
\includegraphics[width=\textwidth]{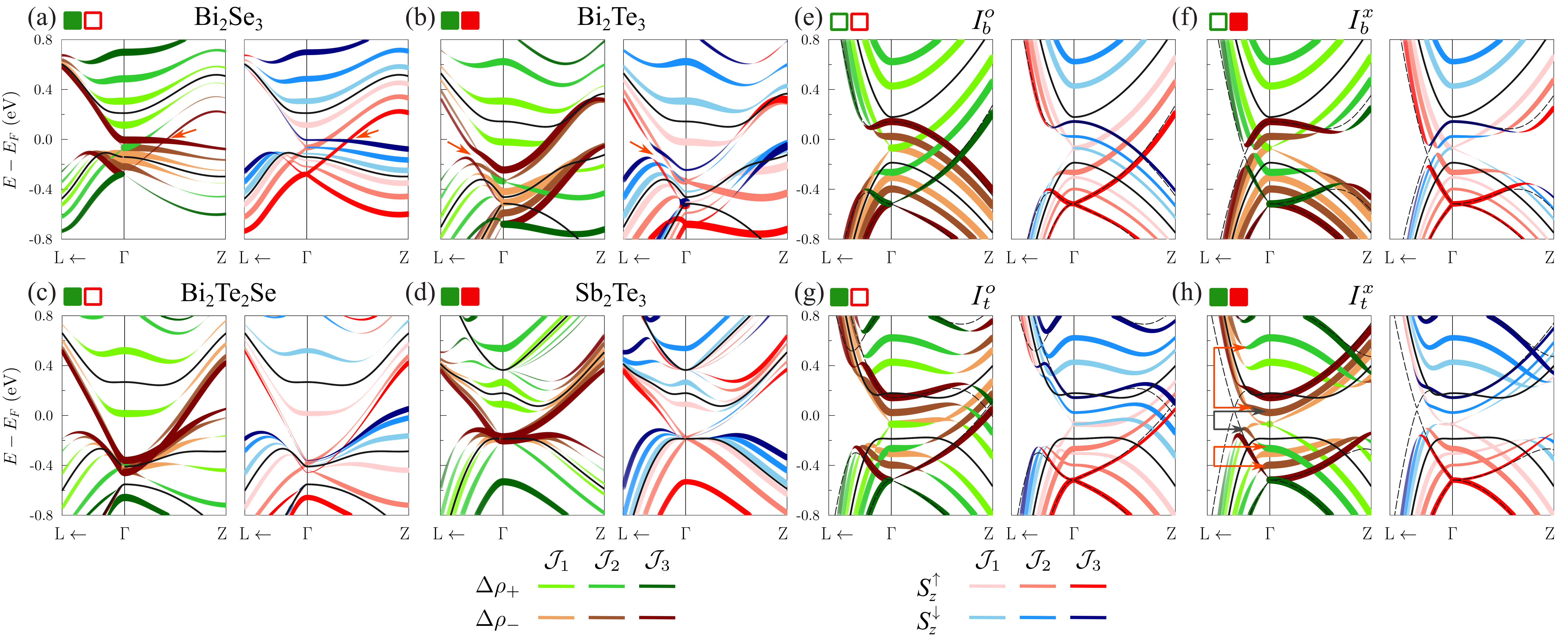}
\caption{(a)-(d) Band structure of the bulk Bi$_2$Se$_3$, Bi$_2$Te$_3$,  Bi$_2$Te$_2$Se and Sb$_2$Te$_3$ by an eight-band $\mathbf{k}\cdot\mathbf{p}$ Hamiltonian at different values of the exchange interaction parameter $\mathcal{J}$ accounting for the appearance of external exchange field along $z$ axis, see $\mathcal{J}_1$, $\mathcal{J}_2$, and $\mathcal{J}_3$ listed in Table~\ref{tab:table_jex}. (e)-(h) Band structure of the bulk model insulators $I_b^{o}$, $I_t^{o}$, $I_b^{x}$, and $I_t^{x}$ (see text) by the four-band Hamiltonian~(\ref{kpHamForBulk}) for $\mathcal{J}_1= \mathcal{J}_c^{\uparrow\downarrow} = 280$~meV, $\mathcal{J}_2= 500$~meV, and $\mathcal{J}_3 = \mathcal{J}^{\uparrow\uparrow}_c=784$~meV. Dashed black lines show the band structure for $\mathcal{J}_3$ by the four-band Hamiltonian~(\ref{kpHamForBulk}) without the third-order terms. In all graphs, black lines show the zero-field band structure. The band inversion is shown by green ($\Delta\rho>0$) and brown ($\Delta\rho<0$) fat bands, see text. Red (blue) fat bands show $S_z>0$ ($S_z<0$) states of the respective model. Green open (filled) squares mark the topologically trivial, $\mathbb{Z}_2=0$ (non-trivial, $\mathbb{Z}_2=1$), zero-field ground state, while red open (filled) squares highlight the state with $\sigma\!\upharpoonleft\!\upharpoonright\!\mu$ ($\sigma\!\downharpoonleft\!\upharpoonright\!\mu$). }
\label{fig5}
\end{figure*}

We now turn to the effect of the external exchange field along $z$ axis on the bulk band structure of the TIs. We will explore both the inversion of the bands that form the edges of the gap and their spin $z$ polarization depending on the exchange parameter $\mathcal{J}$. We put a special focus on the behavior of the band edges at the ``critical'' values of $\mathcal{J}$ at which the conduction and valence bands touch at $\Gamma$. The diagonal form of the spin $z$ component of the spin matrix~(\ref{kpSpinBulk}) simplifies the effect of the exchange term $H_{\rtm{EX}}$, reducing it to a shift of the energies of the basis states $|\Psi_{1\mu}\rangle$ and $|\Psi_{2\mu}\rangle$ at $\Gamma$. Therefore, the critical values are
\begin{align*}
  \mathcal{J}_c^{\uparrow\downarrow} &=(\epsilon_2-\epsilon_1)/(|s^z_1|+|s^z_2|), \\
  \mathcal{J}_c^{\uparrow\uparrow} &=(\epsilon_2-\epsilon_1)/(|s^z_2|-|s^z_1|) \text{ for } |s^z_2|>|s^z_1|,\\
  \mathcal{J}_c^{\downarrow\downarrow} & = (\epsilon_2-\epsilon_1)/(|s^z_1|-|s^z_2|)\text{ for } |s^z_2|<|s^z_1|,
\end{align*}
where the up and down arrows stand for the sign of the expectation spin $z$ projection in the state $|\Psi_{n\mu}\rangle$. Note that the critical values $\mathcal{J}_c^{\uparrow\uparrow}$ and $\mathcal{J}_c^{\downarrow\downarrow}$ are meaningless in the effective models in which the basis functions are treated as eigenfunctions of the spin operator ($s^{z}_{1,2}=1$). In our $\mathbf{k}\cdot\mathbf{p}$ theory, not only the magnitude of $s^{z}_{1,2}$ (which is less than unity) but also its sign is different in different TIs, whereby the magnetic response of each TI acquires individual character. The magnitude of $s^{z}$ determines how fast the respective energy changes with increasing the field, and its sign determines which states (with the opposite or same sign of $J^z$) will come close in energy and interact and whether the valence and conduction bands will cross or anticross. In other words, which matrix elements of the $\mathbf{k}\cdot\mathbf{p}$ Hamiltonian are responsible for the formation of the valence and conduction bands near the Fermi level in the magnetic system. In this regard, we distinguish two cases: the spin $s^z$ and the on-site total angular momentum $J^z$ in the state $|\Psi_{1}\rangle$ have the same sign ($\sigma\!\upharpoonleft\!\upharpoonright\!\mu$) or opposite sign ($\sigma\!\downharpoonleft\!\upharpoonright\!\mu$).

In order to elucidate the effect of the exchange field on the bulk band structure of all the TIs, we derive an eight-band third-order $\mathbf{k}\cdot\mathbf{p}$ Hamiltonian. The resulting bands for the three values of $\mathcal{J}$ listed in Table~\ref{tab:table_jex} are shown in Figs.~\ref{fig5}(a)-\ref{fig5}(d). To illustrate the band inversion, let us consider the contributions $|C_{1(2)}|^2$ of the basis states $|\Psi_{1(2)}\rangle$ to the eigenstates of the magnetic $\mathbf{k}\cdot\mathbf{p}$ Hamiltonian. In Fig.~\ref{fig5}, the fat bands highlighting the overbalance $\Delta\rho=|C_{2}|^2-|C_{1}|^2$ of one of the contributions $|C_{1(2)}|^2$ are green for positive $\Delta\rho$ and brown for negative $\Delta\rho$. Here, we are interested in the effect of the $z$ directed exchange field on the band gap both in the $(k_x,k_y)$ plane at $k_z=0$ and along $k_z$. To this end, we calculate the band structure along the $\Gamma$-$Z$ line, which lies on $z$ axis, and in the $\Gamma$-$L$ direction, which has only a rather small projection on $z$ axis. Hereafter, the Fermi energy is taken to be independent of $\mathcal{J}$.

\begin{table}
\caption{\label{tab:table_jex} Exchange interaction parameter $\mathcal{J}$ (in meV) used in the calculations of bulk band structure, Figs.~\ref{fig5}(a)-\ref{fig5}(d). The asterisk (*) and double asterisk (**) symbols mean $\mathcal{J}_2=\mathcal{J}_c^{\uparrow\downarrow}$ and $\mathcal{J}_3=\mathcal{J}^{\uparrow\uparrow}_c$,   respectively.}
\begin{ruledtabular}
\begin{tabular}{llll}
                              & $\mathcal{J}_1$          & $\mathcal{J}_2$        & $\mathcal{J}_3$      \\
\hline
Bi$_2$Se$_3$                    & 100  &  286$^{\ast}$ & 507$^{\ast\ast}$  \\
Bi$_2$Te$_3$                     & 200 & 584$^{\ast}$   &  1003$^{\ast\ast}$   \\
Bi$_2$Te$_2$Se                 & 300 & 765$^{\ast}$   & 1100    \\
Sb$_2$Te$_3$                    &  100 &  403$^{\ast}$  &   800
\end{tabular}
\end{ruledtabular}
\end{table}

As seen in Figs.~\ref{fig5}(a)-\ref{fig5}(d), the field causes a splitting of the doubly degenerate bands, which is very different for VB and CB due to the very different magnitudes of $s^z_1$ and $s^z_2$, Fig.~\ref{fig2}(a). With increasing the field, the splitting leads to a decrease of the band gap. At $\mathcal{J}=\mathcal{J}_1$, which is below the critical value $\mathcal{J}_c^{\uparrow\downarrow}$, the gap is still open and inverted. The gap is definitely closed at $\mathcal{J}_2 = \mathcal{J}_c^{\uparrow \downarrow}$ in all the TIs: a split-off CB subband touches one of the VB subbands at $\Gamma$. Note that in Bi$_2$Se$_3$ and Bi$_2$Te$_2$Se ($\sigma\!\upharpoonleft\!\upharpoonright\!\mu$) the touch implies that the states $|\Psi_{2\uparrow}\rangle$ and $|\Psi_{1\downarrow}\rangle$ in the presence of the field have equal energies: $\epsilon_1+s_1^z\mathcal{J}=\epsilon_2-s_2^z\mathcal{J}$. In Bi$_2$Te$_3$ and Sb$_2$Te$_3$ ($\sigma\!\downharpoonleft\!\upharpoonright\!\mu$), equal are the field-perturbed energies of $|\Psi_{2\uparrow}\rangle$ and $|\Psi_{1\uparrow}\rangle$: $\epsilon_1-s_1^z\mathcal{J} =\epsilon_2 -s_2^z\mathcal{J}$. In the TIs with $\sigma\!\upharpoonleft\!\upharpoonright\!\mu$ a further increase of the field is accompanied by the crossing of the CB and VB subbands away from $\Gamma$ in the $\Gamma$-$Z$ direction. Thereby a reopening of the gap is prevented, see the orange arrow in Figs.~\ref{fig5}(a). The crossing that occurs at some $k_z$ gives rise to a Weyl node with its partner at $-k_z$. In the TIs with $\sigma\!\downharpoonleft\!\upharpoonright\!\mu$, there is no crossing, and the band structure along $\Gamma L$ and $\Gamma Z$ becomes semimetal-like, which is most clearly seen in Bi$_2$Te$_3$ [the orange arrow in Fig.~\ref{fig5}(b)]. In Sb$_2$Te$_3$, this behavior is less pronounced due to a very small avoided-crossing gap between the subbands in the $\Gamma$-$L$ line.

When reaching the next critical value $\mathcal{J}_c^{\uparrow \downarrow}$ (which is $\mathcal{J}_3$ for Bi$_2$Se$_3$ and Bi$_2$Te$_3$), the field causes one more touch: the CB subband meets the lowest VB subband at $\Gamma$,  Fig.~\ref{fig5}(a) and \ref{fig5}(b). For Sb$_2$Te$_3$ and Bi$_2$Te$_2$Se the value $\mathcal{J}_c^{\uparrow\uparrow}$ is only slightly larger than $\mathcal{J}_c^{\uparrow\downarrow}$, since these TIs have a very small $|s^z_1|$. The spectra of these TIs for $\mathcal{J}_3 > \mathcal{J}_c^{\uparrow\uparrow}$, Fig.~\ref{fig5}(c) and \ref{fig5}(d), demonstrate how the lowest VB subband turns into a non-inverted band that is no longer involved in the properties of the TI spectra near the Fermi level.

The above brief analysis already reveals that owing to the different $s^z$, the conduction and the valence states respond differently to the out-of-plane magnetization, and, depending on the relative sign of $s^z$ and $J^z$ in these states, two distinct scenarios of the magnetization effect arise. In the $\sigma\!\upharpoonleft\!\upharpoonright\!\mu$ TIs (the conventional response), we observe the modifications of the subbands by the field, which in $\Gamma$-$Z$ expectedly result in the appearance of the Weyl nodes along the $k_z$ axis. Along $\Gamma$-$L$ or, more generally, in the $k_z=0$ plane, the modifications manifest themselves in closing the gap and its subsequent reopening similar to the well-studied case of thin films.  The $\sigma\!\downharpoonleft\!\upharpoonright\!\mu$ TIs show a rather complex (unconventional) response to the field both along $\Gamma$-$Z$ and along $\Gamma$-$L$, which has hitherto been overlooked for the topological insulators. In Sec.~\ref{Scenar_Sec}, we present a more detailed analysis of the mentioned scenarios, including the MEI effect on the basis-state composition ($\Delta\rho$) of the subbands and their spin $z$ polarization at different $\mathcal{J}$ for the $\sigma\!\upharpoonleft\!\upharpoonright\!\mu$ and $\sigma\!\downharpoonleft\!\upharpoonright\!\mu$ cases.

\section{Minimal $\mathbf{k}\cdot\mathbf{p}$ model for magnetic system}

\subsection{Four-band representation}

The minimal $\mathbf{k}\cdot\mathbf{p}$ model of the TI bulk crystals has the basis limited to the states $|\Psi_{1}\rangle$ (VB) and $|\Psi_{2}\rangle$ (CB). With this basis set, the microscopically derived $4\times4$ $\mathbf{k}\cdot\mathbf{p}$ Hamiltonian reads
\begin{equation}\label{kpHamForBulk}
H_{\rtm{\mathbf{kp}}}=\left(
\begin{array}{cc}
H_1 &  H_{\rtm{R}} + H_{\spr} + H_{z}  \\
H_{\rtm{R}}^{\dagger}+H_{\spr}^{\dagger}+H_{z}^{\dagger} &  H_2
\end{array}
\right),
\end{equation}
where $H_n=[\epsilon_n+M_{n}^{\spr}k_{\spr}^2 + M_{n}^{z}k_z^2]\rtm{\mathbb{I}}_{2\times2}$ with $k_{\spr}=\sqrt{k_x^2+k_y^2}$. The coupling between the Kramers pairs is described by the following terms:
\begin{eqnarray}
H_{\rtm{R}}&=&\left(
\begin{array}{cc}
-i\theta (k_+^3-k_-^3) & i\alpha_{\spr}k_- \\
-i\alpha_{\spr}k_+        & i\theta(k_+^3-k_-^3)
\end{array}
\right),  \label{Ham_rash}\\
H_{z}&=&\left(
\begin{array}{cc}
 i\beta_{z}k_z         &  i\delta k_+^2k_z \\
-i\delta k_-^2k_z   &  i\beta_{z}k_z
\end{array}
\right),\label{Ham_zet}
\end{eqnarray}
and $H_{\spr}=i\eta(k_+^3+k_-^3)\rtm{\mathbb{I}}_{2\times2}$. Here, $k_{\pm}=k_x\pm ik_y$ and
\begin{eqnarray*}
\alpha_{\spr}&=&\alpha^{(1)}+\alpha^{(3)}k_{\spr}^2+\tilde{\alpha}^{(3)}k_z^2, \\
\beta_{z}&=&\beta^{(1)}+\beta^{(3)}k_{z}^2+\tilde{\beta}^{(3)}k_{\spr}^2.
\end{eqnarray*}
In this basis, the spin matrix is just the $4\times4$ block~(\ref{kpSpinBulk}) of Sec.~\ref{Comp_Det}. The microscopically obtained parameters of the present four-band $\mathbf{k}\cdot\mathbf{p}$ model are listed in Table~\ref{tab:table2} of Appendix~\ref{B_parameters}. Note that the fully {\it ab initio} derived Hamiltonian~(\ref{kpHamForBulk}) is an improvement over the third-order Hamiltonian of Ref.~\cite{Liu_PRB_2010} because it includes additional  third-order terms proportional to $\tilde{\alpha}^{(3)}$, $\tilde{\beta}^{(3)}$, and $\delta$ accounting for the interplay between the in-plane ($k_x$ and $k_y$) and out-of-plane ($k_z$) components. The values of these parameters obtained from the {\it ab initio} spinor wave functions turn out too large to consider these terms negligible. The additional cubic terms describe more accurately the VB and CB extrema located in the mirror plane [shaded areas in Fig.~\ref{fig3}(b)]. In the presence of the $k_z$-directed exchange field, these terms are responsible for the appearance of additional Weyl nodes both in the mirror planes and along the $k_z$ axis. Thus, our four-band $\mathbf{k}\cdot\mathbf{p}$ model fully takes into account the $\mathbf{k}$-cubic effect on the bulk band structure.

Similar to Ref.~\cite{Nechaev_PRBR_2016}, the four-band Hamiltonian~(\ref{kpHamForBulk}) with the parameters listed in Table~\ref{tab:table2} of Appendix~\ref{B_parameters} produces a band gap between the VB and CB in the TI spectra, and for Bi$_2$Se$_3$  and Sb$_2$Te$_3$ near the $\Gamma$ point the width of the gap and the dispersion of its edges are close to that in the original spectra. However, for Bi$_2$Te$_2$Se and Bi$_2$Te$_3$ the $\mathbf{k}\cdot\mathbf{p}$ band gap is substantially narrower than the actual one, and the dispersion of the valence and conduction bands around $\Gamma$ are rather far from the true all-electron bands. For these TIs, a much more accurate description is provided by the eight-band Hamiltonian used in Sec.~\ref{bulkEF_sec}. In this section, we rely on the minimal $\mathbf{k}\cdot\mathbf{p}$ model to reveal the essential physics behind the exchange-field effect on the bulk TI spectra presented in Figs.~\ref{fig5}(a)-\ref{fig5}(d). To this end, in the next subsection we introduce model insulators and apply to them the four-band Hamiltonian~(\ref{kpHamForBulk}).

\subsection{\label{Scenar_Sec}Scenarios of magnetic splitting}

To simplify the analysis of the exchange filed effect on the TI spectra shown in Figs.~\ref{fig5}(a)-\ref{fig5}(d) and, more generally, to outline possible scenarios that can be realized in bulk crystals of magnetic insulators, let us introduce \textit{model} insulators. We start with the Hamiltonian parameters of Sb$_2$Te$_3$ listed in Table~\ref{tab:table2} and only set the effective mass to $M_2^z=-1.89$ and enlarge the spin parameter: $|s^z_1|=0.42$. This gives us two topologically non-trivial model insulators $I_t^{o}$ and  $I_t^{x}$, where the superscripts $o$ and $x$ indicate the ordinary ($s^z_1>0$) and exceptional ($s^z_1<0$) relation between $s^z$ and $J^z$, i.e., the cases $\sigma\!\upharpoonleft\!\upharpoonright\!\mu$ and $\sigma\!\downharpoonleft\!\upharpoonright\!\mu$, respectively. The effective mass $M_2^z$ is modified such as to reduce the CB and VB dispersion curvature along $\Gamma Z$ and thereby facilitate the analysis of the inversion and spin polarization of the bands along $k_z$. However, it also reduces to 0.02~eV the fundamental inverted band gap of $I_t^{o}$ and  $I_t^{x}$, which is located off the high-symmetry lines in the mirror plane. [Note that these topological model insulators are close to being Dirac semimetals with four-fold degenerate Dirac points in the mirror plane: when the effective mass is slightly changed to $M_2^z=-2.59$, there appear points in the mirror plane at which (i)~the diagonal elements $H_1$ and $H_2$ of Eq.~(\ref{kpHamForBulk}) become equal and (ii)~the non-diagonal elements of Eq.~(\ref{kpHamForBulk}) vanish, i.e., both equations $i\alpha_{\spr}k_-+i\delta k_+^2k_z=0$ and $i\beta_zk_z+i\eta(k_+^3+k_-^3)=0$ hold.] By inverting the sign of the effective mass: $M_n^{\spr,z}\rightarrow -M_n^{\spr,z}$, from $I_t^{o}$ and  $I_t^{x}$ we obtain two  trivial (band) insulators $I_b^{o}$ and  $I_b^{x}$. The resulting fundamental trivial band gap of $I_b^{o}$ and  $I_b^{x}$ is $0.37$~eV at $\Gamma$.

Figures~\ref{fig5}(e)-\ref{fig5}(h) show the band structure of the four model insulators for three values of the exchange parameter $\mathcal{J}$. In order to disentangle the band-dispersion curves, let us group the four subbands in two pairs, $P_{\rtm{in}}$ and $P_{\rtm{out}}$, according to their behavior upon the application of the field. (For $I_b^{o}$ and  $I_b^{x}$, along the $\Gamma$-$L$ line these pairs are similar, respectively, to the NS and TS pairs of thin films, which we introduced in Sec.~\ref{Intro}.) Each pair contains one subband of the original doubly degenerate VB and one of the CB. The subbands of the pair $P_{\rtm{in}}$ ($P_{\rtm{out}}$) move towards (away from) each other when the exchange field is introduced. First, we consider the $I_b^{o}$ model insulator, which embodies the $\sigma\!\upharpoonleft\!\upharpoonright\!\mu$ case. Here, the two subbands forming the pair ($P_{\rtm{in}}$ or $P_{\rtm{out}}$) are related to the basis states with different $\mu$. To get insight into the interplay within the pair $P_{\rtm{in}}$, we focus on the related $2\times2$ blocks of the Hamiltonian~(\ref{kpHamForBulk}) and the exchange term $H_{\rtm{EX}}$:
\begin{equation}\label{ordinary_close}
H_{\sigma\!\upharpoonleft\!\upharpoonright\!\mu}=\left(
\begin{array}{cc}
\tilde{\epsilon}_1+s_1^z\mathcal{J} & i\alpha_{\spr}k_-+i\delta k_+^2k_z \\
-i\alpha_{\spr}k_+-i\delta k_-^2k_z   & \tilde{\epsilon}_2-s_2^z\mathcal{J}
\end{array}
\right),
\end{equation}
where $\tilde{\epsilon}_n=\epsilon_n+M_n^{\spr}k_{\spr}^2+M_n^zk_z^2$.

For a finite $\mathcal{J} < \mathcal{J}_c^{\uparrow\downarrow}$ [not shown in Fig.~\ref{fig5}(e)], by construction, the $P_{\rtm{in}}$-gap narrows, while the $P_{\rtm{out}}$-gap gets wider. Both these gaps remain trivial. Next, as seen in Fig.~\ref{fig5}(e), for $\mathcal{J}_1=\mathcal{J}_c^{\uparrow\downarrow}$ the $P_{\rtm{in}}$ gap is closed at $\Gamma$. According to $H_{\sigma\!\upharpoonleft\!\upharpoonright\!\mu}$ of Eq.~(\ref{ordinary_close}), along $\Gamma$-$L$ the leading term of the interaction within the $P_{\rtm{in}}$ pair is linear in $k_{\pm}$, and, therefore, the dispersion around $\Gamma$ is Dirac-like. Beyond this critical point, for $\mathcal{J}_c^{\uparrow\downarrow} < \mathcal{J}_2 < \mathcal{J}_c^{\uparrow\uparrow}$, only the $P_{\rtm{in}}$ gap changes character: it becomes inverted in the $\Gamma$-$L$ direction. Along the $\Gamma$-$Z$ direction, $H_{\sigma\!\upharpoonleft\!\upharpoonright\!\mu}$ has only diagonal terms, and the $P_{\rtm{in}}$ subbands simply cross at some $k_z$. As a consequence, we have a single pair of Weyl nodes on the $k_z$-axis. Upon a further increase of the field up to $\mathcal{J}_3 = \mathcal{J}_c^{\uparrow\uparrow}$ the subbands with the same $\mu$ of different pairs touch at $\Gamma$. Since the interaction of these subbands is linear in $k_{z}$ [see the Hamiltonian~(\ref{kpHamForBulk}) and Eq.~(\ref{extraordinary_close}) below], it is accompanied by a Dirac-like dispersion around $\Gamma$ along the $\Gamma$-$Z$ line.

In $I_{b}^{o}$, the spin $z$ polarization of the subbands by the field in the $\Gamma$-$Z$ direction is trivial, i.e., for every subband the spin $z$ projection $S_z$ has the same sign over the whole $k_z$ interval. Along the $\Gamma$-$L$ line, the polarization of the $P_{\rtm{out}}$ subbands is also trivial, while in the $P_{\rtm{in}}$ pair the subbands are inverted and the polarization is reversed at some point in the $\Gamma$-$L$ line.

In the $I_b^{x}$ model insulator, which differs from $I_b^{o}$ by the sign of $s^z_1$ (the $\sigma\!\downharpoonleft\!\upharpoonright\!\mu$ case), the two subbands of the pair $P_{\rtm{in}}$ or $P_{\rtm{out}}$ originate from the basis states with the same $\mu$, and the $2\times2$ blocks of the Hamiltonian~(\ref{kpHamForBulk}) and $H_{\rtm{EX}}$ that correspond to the $P_{\rtm{in}}$ give the term
\begin{equation}\label{extraordinary_close}
H_{\sigma\!\downharpoonleft\!\upharpoonright\!\mu}=\left(
\begin{array}{cc}
\tilde{\epsilon}_1-s_1^z\mathcal{J} & i\beta_{z}k_z + \mathcal{R}_3\\
 -i\beta_{z}k_z+\mathcal{R}_3^{\ast}   & \tilde{\epsilon}_2-s_2^z\mathcal{J}
\end{array}
\right)
\end{equation}
with $\mathcal{R}_3=i(\eta+\theta)k_+^3+i(\eta-\theta)k_-^3$.

In comparison with $I_b^{o}$, the situation is mirrored: in $I_{b}^{x}$ along the $\Gamma$-$Z$ line the pairs $P_{\rtm{in}}$ and $P_{\rtm{out}}$ behave like the pairs of $I_{b}^{o}$ in the $\Gamma$-$L$ direction with increasing the filed, cf. Fig.~\ref{fig5}(e) and \ref{fig5}(f), also regarding the basis-state composition and the spin $z$ polarization. However, there is an important difference: in the $\Gamma$-$L$ direction the subbands of $P_{\rtm{in}}$, which are expected to cross, similar to the respective subbands of $I_{b}^{o}$ in $\Gamma$-$Z$, anticross due to the $\mathcal{R}_3$ terms of $H_{\sigma\! \downharpoonleft\! \upharpoonright\!\mu}$. As a consequence, the inverted gap reopens for $\mathcal{J} > \mathcal{J}_c^{\uparrow\downarrow}$ in both the $\Gamma$-$L$ and $\Gamma$-$Z$ direction. Due to the avoided crossing between the $P_{\rtm{in}}$ subbands, their spin $z$ polarization reverses sign in the $\Gamma$-$L$ direction as well. Note that the band structure by the second-order Hamiltonian, which neglects the $k_{\pm}^3$ terms [see dashed lines in Figs.~\ref{fig5}(f)], also has a tiny gap along $\Gamma$-$L$, since this line does not lie exactly in the $k_z=0$ plane, but at the same time there is a crossing of the subbands exactly in this plane at all polar angles of $\mathbf{k}_{\spr}$.

We now turn to the $I_t^{o}$ model insulator---the $\sigma\! \upharpoonleft\! \upharpoonright\!\mu$ TI, which derives from $I_b^{o}$ by reversing the sign of the effective masses, so that the pairs $P_{\rtm{in}}$ and $P_{\rtm{out}}$ of $I_t^{o}$ have inverted gaps for $\mathcal{J}$ below $\mathcal{J}_c^{\uparrow\downarrow}$. In the $2\times2$ representation of the $P_{\rtm{in}}$ pair by the term $H_{\sigma\! \upharpoonleft\! \upharpoonright\!\mu}$ of Eq.~(\ref{ordinary_close}), the inversion of its subbands along $\Gamma$-$Z$ can be taken into account by the substitution $\tilde{\epsilon}_1\rightarrow \tilde{\epsilon}_1-\mathcal{M}(\mathbf{k})$ and $\tilde{\epsilon}_2\rightarrow \tilde{\epsilon}_2+\mathcal{M}(\mathbf{k})$, where the correction
\begin{equation*}\label{M_diag}
\mathcal{M}(\mathbf{k})=\frac{\beta^{(1)}k_z^2} {\epsilon_2-\epsilon_1} \left[\beta^{(1)} + 2\beta^{(3)}k_z^2
+\frac{\beta^{(1)}k_z^2} {\epsilon_2-\epsilon_1}(M_1^{z}-M_2^{z})\right]
\end{equation*}
is found by means of the L\"{o}wdin partitioning~\cite{Leowdin_JCP_1951, Schrieffer_PR_1966, Winkler_KP} of the Hamiltonian~(\ref{kpHamForBulk}), retaining terms up to fourth order in $\mathbf{k}$. For $\mathcal{J}=0$, the $\mathcal{M}$-corrected diagonal terms of $H_{\sigma\! \upharpoonleft\! \upharpoonright\!\mu}$ reproduce the zero-field behavior of the CB and VB dispersion [shown in Fig.~\ref{fig5}(g) by black lines] along the $\Gamma$-$Z$ line. Thus, as well as in $I_b^{o}$, for a finite $\mathcal{J}$ it is the field that causes the crossing of the $P_{\rtm{in}}$ subbands along $\Gamma$-$Z$, giving rise to the Weyl nodes on the $k_z$-axis. Also, similar to $I_b^{o}$,  the spin $z$ polarization of the subbands is trivial, despite the changes in their basis-state composition originated from the initial (at $\mathcal{J}=0$) inversion of the $P_{\rtm{in}}$ and $P_{\rtm{out}}$ gaps, see Fig.~\ref{fig5}(g).

In the $\Gamma$-$L$ direction, at $\mathcal{J}=\mathcal{J}_c^{\uparrow \downarrow}$ the $P_{\rtm{in}}$ subbands touch at $\Gamma$ and demonstrate the Dirac-like dispersion. For larger $\mathcal{J}$,  the $P_{\rtm{in}}$ gap is no longer inverted in this direction, while the $P_{\rtm{out}}$ gap preserves its original character, i.e., remains inverted. As in $I_b^{o}$, the inversion of the spin $S_z$ of the subbands accompanies the subband inversion. [Not to be confused with the avoided crossing between the subbands of different pairs at energies from $\sim0.4$ to $\sim0.7$~eV, which is absent in the second-order $\mathbf{k}\cdot\mathbf{p}$ calculations shown by the dashed lines in Fig.~\ref{fig5}(g).]

\begin{figure*}[tbp]
\centering
\includegraphics[width=\textwidth]{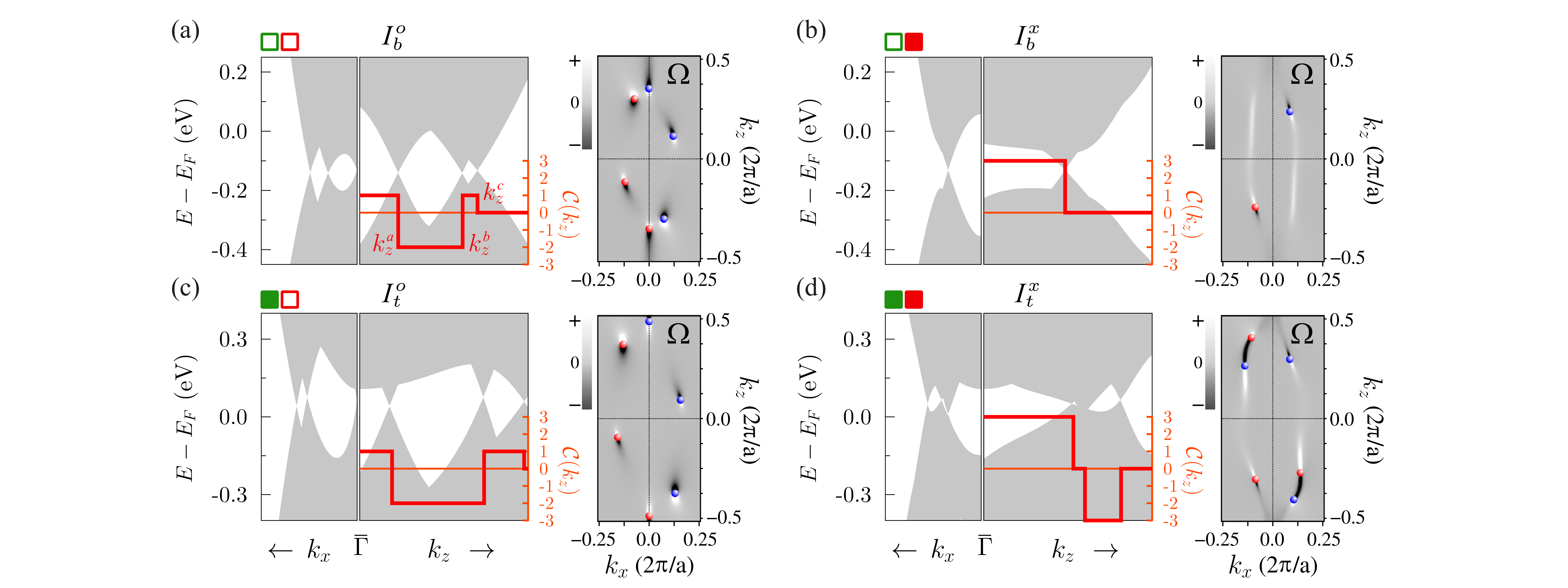}
\caption{Band structure and Berry curvature distribution map of  the model insulators $I_b^{o}$, $I_t^{o}$, $I_b^{x}$, and $I_t^{x}$ in the $(k_x,k_z)$ plane at $k_y=0$. The band structure is shown as a projection onto the $k_x$- and $k_z$-axis. The distribution maps corresponds to the sum of $\Omega_{\lambda}$ of Eq.~(\ref{Berry_curv}) over the two lowest bands. The band structure and Berry curvature of $I_t^{o}$ and $I_t^{x}$ are for $\mathcal{J}=0.70$~eV, while $I_b^{o}$ and $I_b^{x}$ the $\mathcal{J}$ are shown for $\mathcal{J}=0.78$~eV. The Chern number $\mathcal{C}(k_z)$ as a function of $k_z$ is also sown by thick red line superimposed on the $k_z$-projection of the band structure. Green and red squares indicate the same properties as in Fig.~\ref{fig5}.}
\label{fig6}
\end{figure*}

Finally, we consider the $I_t^{x}$ model insulator, which is a $\sigma\! \downharpoonleft\! \upharpoonright\! \mu$ TI. Similar to the relation between $I_b^{x}$ and $I_b^{o}$,  in $I_t^{x}$ we have the $\Gamma$-$L\leftrightarrow\Gamma$-$Z$ mirrored (with respect to $I_t^{o}$) behavior. It is also seen in the corrections to $\tilde{\epsilon}_1$ and $\tilde{\epsilon}_2$ of $H_{\sigma\! \downharpoonleft\! \upharpoonright\!\mu}$ of Eq.~(\ref{extraordinary_close}): in the $\Gamma$-$L$ direction (or, more precisely, in the $k_z=0$ plane) the diagonal terms of the latter are $\tilde{\epsilon}_1-\mathcal{N}(\mathbf{k})$ and $\tilde{\epsilon}_2+\mathcal{N}(\mathbf{k})$ with
\begin{equation*}\label{N_diag}
\mathcal{N}(\mathbf{k})=\frac{\alpha^{(1)}k_{\spr}^2} {\epsilon_2-\epsilon_1} \left[\alpha^{(1)}+2\alpha^{(3)}k_{\spr}^2
+\frac{\alpha^{(1)}k_{\spr}^2} {\epsilon_2-\epsilon_1}(M_1^{\spr}-M_2^{\spr})\right].
\end{equation*}
In this direction, for $\mathcal{J}\geqslant \mathcal{J}_c^{\uparrow \downarrow}$ the avoided crossing of the $P_{\rtm{in}}$ subbands described by the corrected $H_{\sigma\! \downharpoonleft\! \upharpoonright\!\mu}$ is due to the cubic term $\mathcal{R}_3$. In the absence of this term, in any $\mathbf{k}_{\spr}$ direction these subbands just cross at some $k_{\spr}$, preserving their basis-state composition reflecting the inversion of the $I_t^{x}$ band gap in the zero-field state, similar to the strong MEI case shown in the pale blue panel of Fig.~\ref{fig1}(b). The cubic term brings about the inversion of the $P_{\rtm{in}}$ subbands, which is accompanied by the sign reversal of the spin $z$ polarization, see Fig.~\ref{fig5}(h). [In Fig.~\ref{fig5}(h), for $\mathcal{J}_2$ the gray arrows mark the avoided-crossing gap between the $P_{\rtm{in}}$ subbands, and the orange arrows indicate the positions of the ``survived'' inverted gaps between the subbands of the TS pairs of Sec.~\ref{Intro}.] Along $\Gamma$-$Z$, the gap of the $P_{\rtm{in}}$ pair becomes trivial for $\mathcal{J}> \mathcal{J}_c^{\uparrow \downarrow}$, and its subbands are trivially polarized.

Let us now come back to the real TIs. In the spectra in Figs.~\ref{fig5}(a)-\ref{fig5}(d), we now can recognize the characteristic features of the paired subbands---their basis-state composition and spin $z$ polarization. In spite of the presence of additional bands due to the larger size of the model basis, one can clearly see the fundamental modifications of the band-gap edges and their polarization by the out-of-plane exchange field, which are revealed by the detailed analysis of the model insulators within the minimal third-order $\mathbf{k}\cdot\mathbf{p}$ model. Regarding the response to the field, the distinction between the real and model TIs lies in the smaller ratio $|s^z_1/s^z_2|$ in real crystals. For Bi$_2$Te$_3$ and Bi$_2$Se$_3$, this ratio is around 0.3, which is comparable with the value of 0.5 for the model TIs, and, therefore, their response to the field resembles that of $I_t^{x}$ and $I_t^{o}$, respectively. For Bi$_2$Te$_2$Se and Sb$_2$Te$_3$, the ratio is smaller than 0.1 and, as a consequence, in these TIs around $\Gamma$ the subbands of the split VB respond rather weakly and even for a field of a moderate strength give a small energy separation between the valence and conduction bands.

\section{\label{Chenum_Sec}Weyl nodes and Chern number}

The above analysis shows that the $z$-directed exchange field gives rise to Weyl nodes only in the $\sigma \!\upharpoonleft \!\upharpoonright\!\mu$ case and along the $k_z$-axis. Let us now analyze the behavior of the VB and CB subbands of the model insulators away from high-symmetry lines. We consider the field-induced changes over the whole mirror plane of the BZ, Fig.~\ref{fig3}, in which lie the CB and VB extrema of the real TIs~\cite{Nechaev_PRB_2013_BISE, Nechaev_PRB_2013_BITE, Nechaev_PRB_2015_SBTE} and of the model insulators $I_t^{o,x}$, so that a touch of the VB and CB subbands may take place in this plane at some value of $\mathcal{J}$.

To characterize the topology of the resulting spectra under the $z$-directed exchange field, we calculate the Chern number as a function of $k_z$ along $\Gamma$-$Z$ from the formula~\cite{Wang_PRB_2006, Weng_AP_2015}
\begin{equation}\label{CHENUM}
\mathcal{C}(k_z) =\frac{1}{2\pi} \sum_{\lambda} \int d\mathbf{k}_{\spr} f_{\lambda}(\mathbf{k}) \Omega_{\lambda}(\mathbf{k}),
\end{equation}
where $f_{\lambda}(\mathbf{k})$ is the Fermi factor, and $\Omega_{\lambda}(\mathbf{k})$ is the Berry curvature
\begin{equation}\label{Berry_curv}
\Omega_{\lambda}(\mathbf{k})=-2\rtm{Im}\sum_{\lambda'\neq\lambda} \frac{\langle \mathbf{C}_{\mathbf{k}}^{\lambda}|V_x |\mathbf{C}_{\mathbf{k}}^{\lambda'}\rangle \langle \mathbf{C}_{\mathbf{k}}^{\lambda'}|V_y |\mathbf{C}_{\mathbf{k}}^{\lambda}\rangle}{(E_{\mathbf{k}}^{\lambda} -E_{\mathbf{k}}^{\lambda'})^2}.
\end{equation}
Here, $V_{x,y}=\partial H_{\rtm{\mathbf{kp}}}/\partial k_{x,y}$, and the vectors $\mathbf{C}_{\mathbf{k}}^{\lambda}$ diagonalize the Hamiltonian $H=H_{\rtm{\mathbf{kp}}} + H_{\rtm{EX}}$, $H|\mathbf{C}_{\mathbf{k}}^{\lambda}\rangle = E_{\mathbf{k}}^{\lambda} |\mathbf{C}_{\mathbf{k}}^{\lambda}\rangle$. We assume an insulating band structure in the ($k_x,k_y$) plane at every $k_z$, so in calculating $\mathcal{C}(k_z)$ the Fermi energy changes with $k_z$ accordingly.

For $\mathcal{J}=0$, all the model insulators are characterized by zero Chern number~(\ref{CHENUM}).  In  $I_b^{o}$ and $I_b^{x}$, a phase with a nonzero $\mathcal{C}(k_z)$ emerges at $\mathcal{J} > \mathcal{J}_c^{\uparrow \downarrow}$. In $I_b^{o}$, right above $\mathcal{J}_1=\mathcal{J}_c^{\uparrow \downarrow}$ a single pair of Weyl nodes with opposite chiral charges ($-1$ and $+1$) appears on the $k_z$-axis (at $k_z^c$ and $-k_z^c$). Then, at a larger $\mathcal{J}$ (less than $\mathcal{J}_2$) the VB and CB subbands touch in the mirror plane, giving rise to Weyl nodes in this plane. For $k_z>0$, in Fig.~\ref{fig3}(b) this occurs in the pale orange parts of the mirror plane, while for $k_z<0$, according to the spatial-inversion symmetry, it takes place in the pale blue parts. Due to the $C_{3v}$ symmetry the mirror-plane nodes appear in triples, and each set of three nodes carries a chiral charge $+3$ or $-3$ (similar sets of three Weyl nodes appear in the ferromagnetic HgCr$_2$Se$_4$ with the magnetization along the [111] axis~\cite{Fang_PRL_2014}).

Hereafter, we restrict ourselves to the ($k_x,k_z$) plane at $k_y=0$, the pale orange and pale blue part of which refer to positive and negative $k_x$, respectively, Fig.~\ref{fig3}(b). In this plane, there are four nodes with $k_xk_z>0$: two oppositely charged nodes with $k_x>0$ and $k_z>0$ and their partners with $k_x<0$ and $k_z<0$. With increasing the field, the positively charged node with $k_z>0$ and negatively charged node with $k_z<0$ move in the plane towards the $k_z$ axis and then pass it through, so $k_x$ changes sign, and now there are two $k_xk_z<0$ nodes. In passing through the axis, these mirror-plane nodes merge with the $k_z$-axis nodes into double-Weyl nodes having charge $\pm2$. This occurs at a $\mathcal{J}$ a few meV above $\mathcal{J}_2$ at which the $P_{\rtm{in}}$ subbands cross along the $\Gamma$-$Z$ line at $k_z^{dW} = \pm\sqrt{-\alpha_1^{(1)}/ \tilde{\alpha}^{(3)}}$: in the non-diagonal element of $H_{\sigma\! \upharpoonleft\! \upharpoonright\!\mu}$ the term $i\alpha_{\spr}k_-$ is reduced to $i\alpha^{(3)}k_{\spr}^2k_-$, and it is the term $i\delta k_{+}^2k_z$ that is responsible for the $k_{\spr}$-quadratic dispersion of the subbands around the crossing point as, e.g., in Refs.~\cite{Onoda_JPSJ_2002, Xu_PRL_2011}. As a result, at $\mathcal{J} \sim \mathcal{J}_c^{\uparrow \uparrow}$ we arrive at the mirror-plane nodes with $\pm k_z^a$ and $\pm k_z^b$ as shown by red and blue balls in Fig.~\ref{fig3}(b).

Figure~\ref{fig6}(a) shows for $\mathcal{J}_3 = \mathcal{J}_c^{\uparrow \uparrow}$ the band structure of $I_b^{o}$ in the $k_y=0$ plane projected onto the $k_x$- and $k_z$-axes and the related Chern number as a function of $k_z$. As clearly seen in the figure, in the $k_x$ direction ($\bar{\Gamma}$-$\bar{M}$) there is a cone at $\bar{\Gamma}$ related to the pair of the Weyl nodes on the $k_z$-axis. Two other cones at finite $k_x$ correspond to the mirror-plane Weyl nodes. As seen in the figure, because of these nodes the Chern number undergoes two jumps in going from $\Gamma$ to $Z$: the first one of $-3$ at $k_z^a$ and the second one of $+3$ at $k_z^b$. From $k_z=0$ to $k_z^a$ and from $k_z^b$ to $k_z^c$, the Chern number $\mathcal{C}(k_z)$ equals 1 and at $k_z^c$ drops to zero. Figure~\ref{fig6}(a) also shows the Berry curvature distribution in the $k_y=0$ plane for $I_b^{o}$ at $\mathcal{J}_3$, which reveals the location of the nodes [depicted by red and blue balls as in Fig.~\ref{fig3}(b)] since the Berry curvature is singular at a Weyl node.

In the $I_b^{x}$ model insulator, there are no Weyl nodes along the $\Gamma$-$Z$ line. Here, the phase with a nonzero $\mathcal{C}(k_z)$ emerges right above $\mathcal{J}_c^{\uparrow \downarrow}$ due to the mirror-plane nodes: the touch of the $P_{\rtm{in}}$ subbands at $\Gamma$ leads to the appearance of a negatively charged Weyl node with $k_x>0$ and $k_z>0$ and its positively charged partner with  $k_x<0$  and $k_z<0$ in the ($k_x,k_z$) plane. A further increase of the field moves the nodes in the mirror plane away from $\Gamma$. The resulting set of the Weyl nodes in the BZ is similar to that in the magnetic Weyl semimetal Co$_3$Sn$_2$S$_2$, which has the same space group as the 3D TIs~\cite{Liu_NatPhys_2018}. Figure~\ref{fig6}(b) shows the projected band structure of $I_b^{x}$, the Chern number $\mathcal{C}(k_z)$, and the Berry curvature distribution for $\mathcal{J}_3 = \mathcal{J}_c^{\uparrow \uparrow}$: the Chern number is $+3$ up to the $k_z$ of the mirror-plane nodes indicated on the Berry curvature distribution map, where $\mathcal{C}(k_z)$ vanishes. This demonstrates that the topology of the resulting spectrum of the insulator under magnetic exchange field changes drastically by reversing the sign of $s_1^z$.

The $I_t^{o}$ and $I_t^{x}$ insulators have a narrow inverted fundamental gap, and, therefore, already at a rather small $\mathcal{J} < \mathcal{J}_c^{\uparrow \downarrow}$ the VB and CB subbands touch in the mirror plane, giving rise to the mirror-plane Weyl nodes in a similar way as in $I_b^{o}$ for $\mathcal{J} > \mathcal{J}_c^{\uparrow \downarrow}$. Moreover, in $I_t^{o}$ an increase of the field (still below $\mathcal{J}_c^{\uparrow \downarrow}$) leads to the second touch of the subbands that creates the mirror-plane nodes with $k_xk_z<0$. As a result, until the $k_z$-axis Weyl nodes appear, there are four sets of three mirror-plane nodes with positive $k_z$ and, consequently, four sets with negative $k_z$. Upon a further increase of the filed, half of these sets annihilate, and right above $\mathcal{J} = \mathcal{J}_c^{\uparrow \downarrow}$ we arrive at a stable configuration of the mirror-plane and $k_z$-axis Weyl nodes similar to that in $I_b^{o}$, cf. Fig.~\ref{fig6}(a) and Fig.~\ref{fig6}(c). As a result, as seen in the figures, the Chern number $\mathcal{C}(k_z)$ and the cone-like features of the projected band structure behave similarly to the case of $I_b^{o}$.

In $I_t^{x}$, the second touch takes place at some $\mathcal{J} > \mathcal{J}_c^{\uparrow \downarrow}$, after half of the mirror-plane nodes produced by the first touch have already annihilated in $\Gamma$ at $\mathcal{J}_c^{\uparrow \downarrow}$. In the projected band structure along $k_z$, Fig.~\ref{fig6}(d), the cone that corresponds to the remaining nodes is located between the cones related to the second-touch nodes. Because of these second-touch nodes, which are clearly distinguished in the Berry curvature distribution map in the $k_xk_z<0$ quadrants of the ($k_x,k_z$) plane, the Chern number has two jumps more than in the $I_b^{x}$ insulator.

\section{\label{thin_film_sec}Thin films}

We focus now on the question of whether the unconventional response to the field may occur in the films of 3D TIs. For the bulk crystals, we have shown that there is a strong difference in the spectrum topology, depending on what type of response is realized. Therefore, we expect the different responses to lead to different QAH states of the films. Let us consider first the thickness dependence of the spin expectation values $s^{\spr}$ and $s^z$ in the valence and conduction states of the films.

\begin{figure}[tbp]
\centering
\includegraphics[width=\columnwidth]{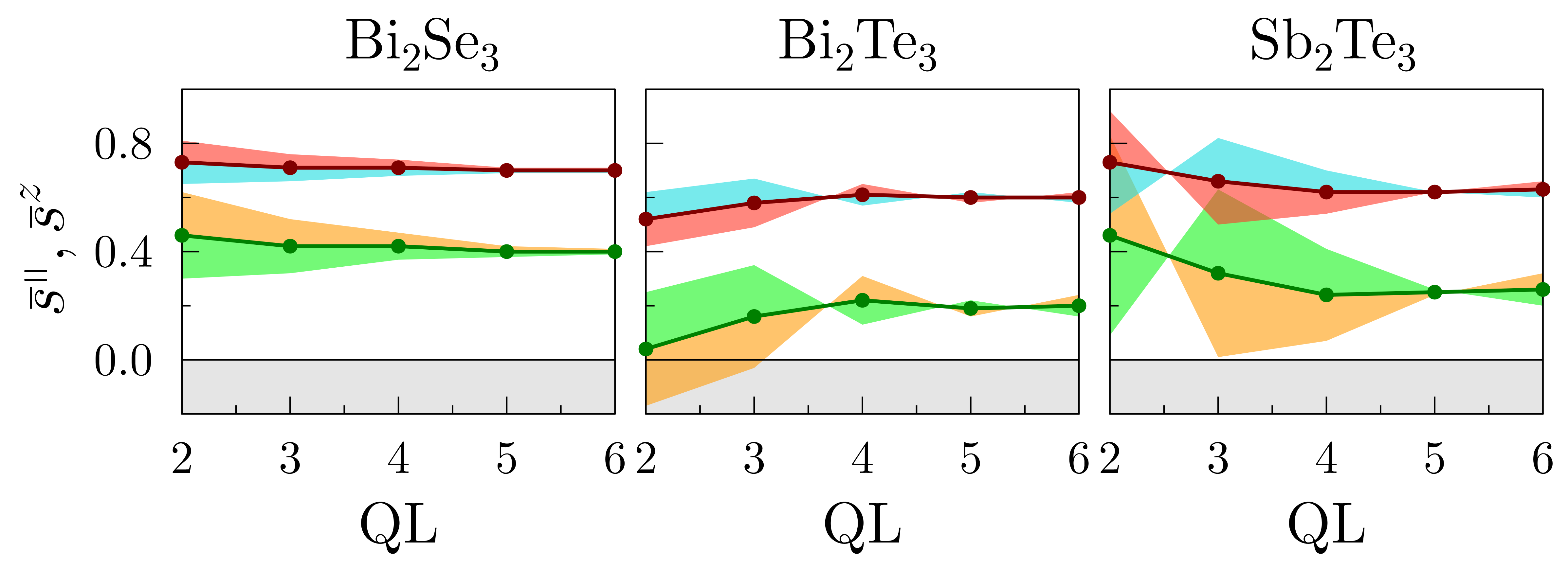}
\caption{In-plane $\bar{s}^{\spr}=\frac{1}{2}(s_1^{\spr}+s_2^{\spr})$ (brown lines and dots) and out-of plane $\bar{s}^z = \frac{1}{2} (s_1^{z}+s_2^{z})$ (green lines and dots) spin as a function of the film thickness. The outer borders of the colored areas show $s^{\spr}_1$ (red), $s^z_1$ (orange), $s^{\spr}_2$ (blue), and $s^z_2$ (green). }
\label{fig7}
\end{figure}

As seen in Fig.~\ref{fig7}, in the Bi$_2$Se$_3$ films, which are trivial two-dimensional (2D) insulators for all the thicknesses considered~\cite{Nechaev_PRBR_2016}, the spin parameters $s^{\spr,z}_1$ and $s^{\spr,z}_2$ converge steadily to the averaged spin $\bar{s}^{\spr,z} = \frac{1}{2} (s^{\spr,z}_1 + s^{\spr,z}_2)$. In contrast, the spins in the films of Bi$_2$Te$_3$ and Sb$_2$Te$_3$ oscillate with the amplitude $\Delta s^{\spr,z} = \frac{1}{2} (s_1^{\spr,z}-s_2^{\spr,z})$ around $\bar{s}^{\spr,z}$ with the number of QLs. (Note that these oscillations do not correlate with the $\mathbb{Z}_2$ topological invariant obtained from the parities of the original wave functions at the TRIMs of the 2D BZ~\cite{Nechaev_PRBR_2016}, i.e., they cannot be associated with the inversion of the bands at $\bar{\Gamma}$.) The amplitude of the oscillations is larger for the out-of-plane spin than for the in-plane spin and decreases with the film thickness. For the films thicker than 4QLs, the oscillations become unimportant, and due to the positive $\bar{s}^{\spr,z}$ the response to an external exchange field should be similar to that in the common models of the Dirac surface state. For the thinner films (from 2QLs to 4QLs), only in two- and three-QL films of Bi$_2$Te$_3$ the oscillations reach negative values of $s^{z}_1$, which implies an unconventional response of the film band-gap edges to the field, which does not fit into the scenarios reported in Ref.~\cite{Yu_Science_2010} [see the strong MEI case in the pale blue panel of Fig.~\ref{fig1}(b)]. The nearly zero values of $s_1^z$ in the three-QL films of Bi$_2$Te$_3$ and Sb$_2$Te$_3$ (negative and positive, respectively, Fig.~\ref{fig7}) cause a very low sensitivity of the valence bands to the field as compared with the conduction band, similar to the bulk crystals Sb$_2$Te$_3$ and Bi$_2$Te$_2$Se.

Next, we construct a minimal $\mathbf{k}\cdot\mathbf{p}$ model for the films. For the centrosymmetric TI films in the periodic slab model we obtain the same Hamiltonian~(\ref{kpHamForBulk}), where $k_z$ is set to zero. The bulk-like representation of the Hamiltonian~(\ref{kpHamForBulk}) can be decomposed into the blocks related to the film surfaces and the blocks accounting for their interaction through the film. To this end, we transfer to the new basis~\cite{Nechaev_PRB_2018, Nechaev_PRB_2019} $|\Phi^{\pm}_{\mu}\rangle =\frac{1}{\sqrt{2}} \left[|\Psi_{1\mu}\rangle \pm |\Psi_{2\mu}\rangle\right]$ in which the original Hamiltonian (\ref{kpHamForBulk}) reads
\begin{equation}\label{HamTwoSurfaces}
H_{\rtm{\mathbf{kp}}}\longrightarrow H^{\rtm{Film}}_{\rtm{\mathbf{kp}}}=\left(
\begin{array}{cc}
H_{\rtm{Surf}}^{+} & H_{\rtm{int}}  \\
H^{\dag}_{\rtm{int}} & H_{\rtm{Surf}}^{-}
\end{array}
\right).
\end{equation}
Here, $H_{\rtm{Surf}}^{\pm}=[\epsilon+M^{\spr}k_{\spr}^2]\rtm{\mathbb{I}}_{2\times2}\pm H_{\rtm{R}}$ and the interaction term $H_{\rtm{int}}=[\Delta\epsilon+\Delta M^{\spr}k_{\spr}^2] \rtm{\mathbb{I}}_{2\times2} - H_{\spr}$, with $k_x$ being the $\bar{\Gamma}$-$\bar{M}$ projection.

In the new basis, the spin matrix~(\ref{kpSpinBulk}) has the form
\begin{equation}\label{SpinTwoSurfaces}
\rtm{\mathbf{S}}^{4\times4}_{\rtm{\mathbf{kp}}}\longrightarrow \rtm{\mathbf{S}}^{\rtm{Film}}_{\rtm{\mathbf{kp}}}=\left(
\begin{array}{cc}
\bar{\rtm{\mathbf{S}}} & \Delta\rtm{\mathbf{S}}  \\
\Delta\rtm{\mathbf{S}} & \bar{\rtm{\mathbf{S}}}
\end{array}
\right),
\end{equation}
with $\bar{\rtm{\mathbf{S}}}=(\bar{s}^{\spr}\bm{\sigma}_{\spr}, \bar{s}^{z}\sigma_z)$ and $\Delta\rtm{\mathbf{S}}=(\Delta s^{\spr}\bm{\sigma}_{\spr}, \Delta s^{z}\sigma_z)$, see Fig.~\ref{fig7}. The parameters in Eqs.~(\ref{HamTwoSurfaces}) and (\ref{SpinTwoSurfaces}) for all the films are listed in Tables~\ref{tab:table3}-\ref{tab:table5} of Appendix~\ref{B_parameters}. From these tables, it is clearly seen how the surface parameters converge with the film thickness, while the interaction terms tend to vanish. Note that starting from two- or three-QL films the Hamiltonian~(\ref{HamTwoSurfaces}) and the spin matrix~(\ref{SpinTwoSurfaces}) can be thought of in terms of a decomposition into two copies of a Dirac electronic system and their interaction, which is described up to third order in $\mathbf{k}$.

\begin{figure}[tbp]
\centering
\includegraphics[width=\columnwidth]{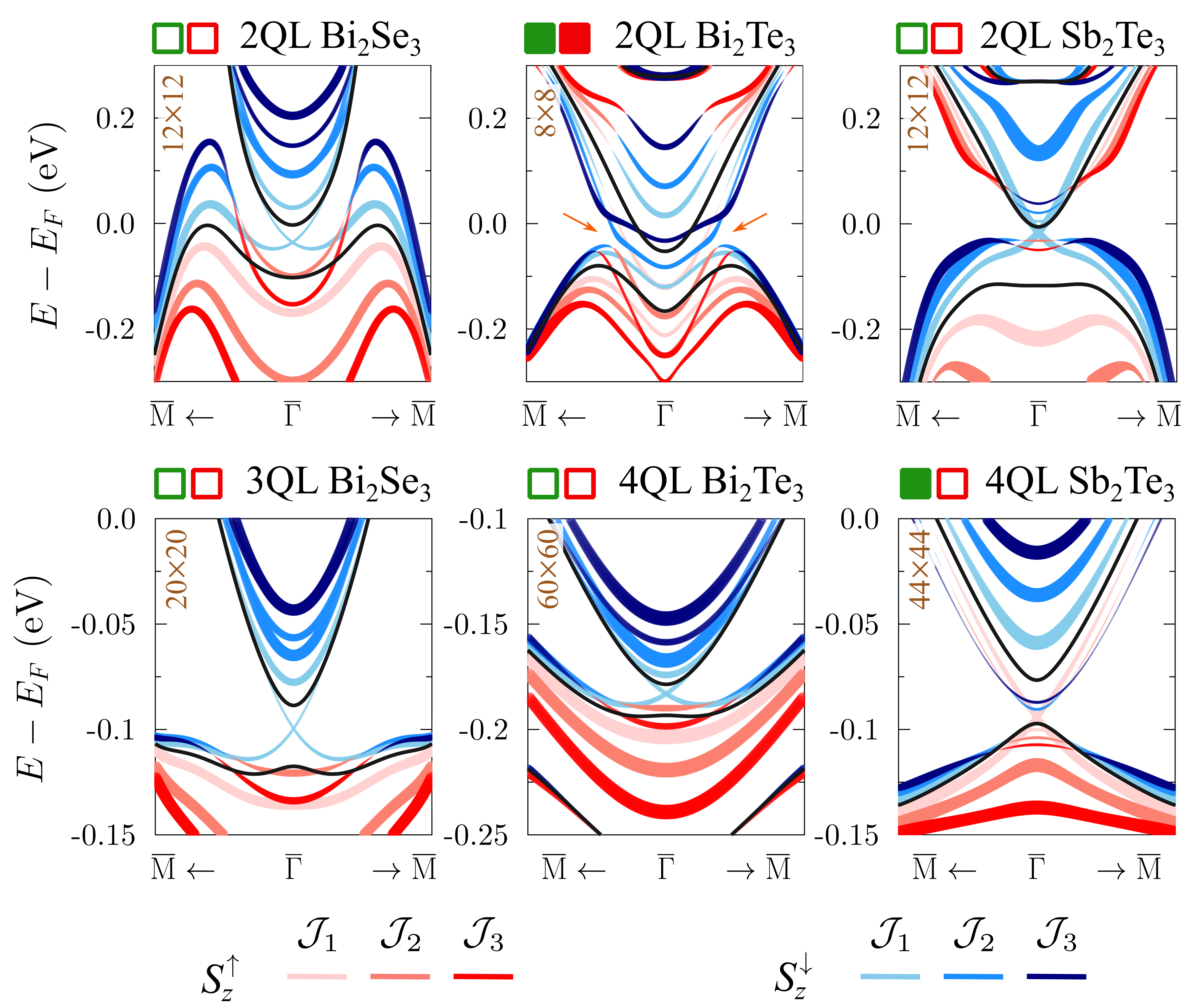}
\caption{Spin-resolved band structure of the TI films from two to four QLs by our $\mathbf{k}\cdot\mathbf{p}$ models of different size at the values of the exchange-interaction parameter $\mathcal{J}$ listed in Table.~\ref{tab:table_jex_Films} (the thickness of the film and the size of the related model are shown in the graphs). Black lines represent the zero-field band structure. Red and blue fat bands reveal the spin $z$ polarization ($S_z>0$ and $S_z<0$, respectively) of the model states. Green and red squares indicate the same properties as in Fig.~\ref{fig5}.}
\label{fig8}
\end{figure}

\begin{table}
\caption{\label{tab:table_jex_Films} Same as in Table~\ref{tab:table_jex}, but for the exchange interaction parameter $\mathcal{J}$ (in meV) used in the band-structure calculations of the thin films, Fig.~\ref{fig8}. The asterisk ($\ast$) and the double star ($\star\star$) symbols mean $\mathcal{J}_1=\mathcal{J}_c^{\uparrow\downarrow}$ and $\mathcal{J}_2$ or $\mathcal{J}_3=\mathcal{J}_c^{\downarrow\downarrow}$, respectively.}
\begin{ruledtabular}
\begin{tabular}{clll}
                              & $\mathcal{J}_1$          & $\mathcal{J}_2$        & $\mathcal{J}_3$      \\
\hline
\multicolumn{4}{l}{\bf Bi$_2$Se$_3$}  \\
\hspace{10pt}2QL                      & 109$^{\ast}$  &  319$^{\star\star}$ & 500  \\
\hspace{10pt}3QL                      & 34$^{\ast}$  &  100 & 141$^{\star\star}$   \\
\multicolumn{4}{l}{\bf Bi$_2$Te$_3$}  \\
\hspace{10pt}2QL                      & 274$^{\ast}$  &  500 & 800  \\
\hspace{10pt}4QL                      & 34$^{\ast}$  &  85$^{\star\star}$ & 150  \\
\multicolumn{4}{l}{\bf Sb$_2$Te$_3$}  \\
\hspace{10pt}2QL                      & 121$^{\ast}$  &  300 & 500  \\
\hspace{10pt}4QL                      & 44$^{\ast}$  &  100 & 150
\end{tabular}
\end{ruledtabular}
\end{table}

The minimal $\mathbf{k}\cdot\mathbf{p}$ model for the centrosymmetric films provides an accurate description of the dispersion of the valence and conduction bands in a close vicinity of the $\bar{\Gamma}$ point, while away from this point the dispersion strongly deviates from the all-electron bands, and for most of the films the model does not produce a band gap. Most important is that owing to the same form of the bulk and film four-band Hamiltonians in the $\left\{\Psi_{1\mu},\Psi_{2\mu}\right\}$ basis, in the case of the TI films the modification of the subbands of the split valence and conduction bands by the exchange field is expected to be similar to that of the $P_{\rtm{in}}$ and $P_{\rtm{out}}$ subbands of the bulk model insulators in the $\Gamma$-$L$ direction, Sec.~\ref{Scenar_Sec}. What is new here is that for some film thicknesses the VB spin parameter is larger than the CB one: $s_1^z > s_2^z$. As a consequence, the scenario of the critical value $\mathcal{J}_c^{\downarrow\downarrow}$ can be realized, in accord with the \textit{ab initio} calculations of Ref.~\cite{Yu_Science_2010}. Thus, based on the similarity between the subbands of the TI films and bulk crystals, we can interpret the results of our $\mathbf{k}\cdot\mathbf{p}$ simulation of the exchange-field effect in the films.

In order to accurately model the response of the TI films to the $z$ directed exchange field, we generate now large-size $\mathbf{k}\cdot\mathbf{p}$ Hamiltonians. The spin-resolved band structure by these Hamiltonians for three values of the exchange-interaction parameter, Table~\ref{tab:table_jex_Films}, is shown in Fig.~\ref{fig8}: at $\mathcal{J}_1 = \mathcal{J}_c^{\uparrow\downarrow}$ the $\bar{\Gamma}$ gap closes, and the two-QL Bi$_2$Te$_3$ ($\sigma\!\downharpoonleft\!\upharpoonright\!\mu$ case) is the only film with a non-Dirac-like dispersion around $\bar{\Gamma}$. It follows from the four-band analysis of Sec.~\ref{Scenar_Sec} that for this topologically non-trivial film (the QSH insulator) the critical value $\mathcal{J}_c^{\uparrow\downarrow}$ of the exchange-interaction parameter does not lift the initial (zero-field) inversion in any of the pairs of the subbands with different $\mu$, i.e., the thin film TS pairs introduced in Sec.~\ref{Intro} [see also the orange arrows in Fig.~\ref{fig5}(h)]. In contrast, other spectra in Fig.~\ref{fig8} demonstrate the $k_{\spr}$-linear behavior as expected for the $\sigma\!\upharpoonleft\!\upharpoonright\!\mu$ case (cf. the \textit{ab initio} calculations of the Bi$_2$Se$_3$ three-QL film in Ref.~\cite{Jin_PRB_2011}). Here, for the films trivial at $\mathcal{J}=0$ [the two-QL film of Bi$_2$Se$_3$  and Sb$_2$Te$_3$, the three-QL film of Bi$_2$Se$_3$, and the four-QL film of Bi$_2$Te$_3$], $\mathcal{J}_c^{\uparrow\downarrow}$ induces the inversion in the $P_{\rtm{in}}$ pair in the $\Gamma$-$L$ direction of $I_b^{o}$. In contrast, for the non-trivial four-QL film of Sb$_2$Se$_3$ (the QSH insulator) it lifts the inversion in the pair, cf. the $\Gamma$-$L$ direction of $I_t^{o}$. Thus, in the $\sigma\!\upharpoonleft\!\upharpoonright\!\mu$ films we have the threshold for a quantum phase transition to a QAH state with the Chern number $\mathcal{C}=+1$ as was described in Ref.~\cite{Yu_Science_2010}. Note that for the films trivial at $\mathcal{J}=0$ in Fig.~\ref{fig8} the spin parameters are related as $s_1^z > s_2^z$, and, therefore, the Dirac-like subbands have a negative spin $z$ polarization, while in $I_b^{o}$, where $s_1^z < s_2^z$, it is positive.

Following the analogy with the bulk model insulators, we note that a further increase of the field in the non-trivial two-QL film of Bi$_2$Te$_3$ causes a reopening of the band gap of the film, which results from the avoided crossing (see orange arrows in Fig.~\ref{fig8}) of the $P_{\rtm{in}}$ subbands (or, in other words, of two subbands belonging to different TS pairs, Fig.~\ref{fig1}(b), both of which are inverted, and one of them becomes trivial at $\mathcal{J}\sim1.4$~eV). Obviously, here we see an unconventional response of the film band-gap edges to the $z$-directed exchange field similar to that observed in the $\Gamma$-$L$ direction of the model insulator $I_t^{x}$, Fig.~\ref{fig5}(h). Here, for the two-QL Bi$_2$Te$_3$ the Chern number $\mathcal{C}$ acquires the value of $-3$, implying a transition from a QSH state to a large-Chern-number QAH state (cf. the Co$_3$Sn$_3$S$_2$ film with $\mathcal{C}=3$ as a 2D limit of the magnetic Weyl semimetal Co$_3$Sn$_2$S$_2$ having only the mirror-plane Weyl nodes~\cite{Muechler_PRB_2020}). In the $\sigma \! \upharpoonleft \! \upharpoonright \! \mu$ films, the further increasing of the field has no effect on the topology of the subbands. In the four-QL film of Sb$_2$Te$_3$, the modification of the spin $z$ polarization of the subbands by the field resembles that in $I_t^{o}$, while owing to the relation $s_1^z > s_2^z$, the $P_{\rtm{in}}$ pair of the initially trivial films shown in Fig.~\ref{fig8} behave differently from the $P_{\rtm{in}}$ pair of $I_b^{o}$.

\section{Conclusions}

To summarize, we applied the \textit{ab initio} relativistic $\mathbf{k}\cdot\mathbf{p}$ theory to studying the effect of external exchange field on the band structure of bulk crystals and thin films of the 3D TIs.  We have focused on the modifications of the valence and conduction bands by the field perpendicular to the QL plane. The effect of the field of different strength is simulated by an exchange term that influences the bands according to their real spin, specifically, the spin $z$ expectation value $s^z$ in the valence and conduction states---the Kramers-degenerate eigenstates of the $\Gamma$-point-projected \textit{ab initio} Hamiltonian. We have shown that the conduction-state $s^z$ is close to unity and varies slightly from one TI to another, while the material's individuality manifests itself in the valence-state $s^z$, the magnitude and even the sign of which is different in different TIs [it is rather large and positive for Bi$_2$Se$_3$, large and negative for Bi$_2$Te$_3$, but weakly positive for Bi$_2$Te$_2$Se and weakly negative for Sb$_2$Te$_3$]. Thus, the present theory reveals the individual character of the magnetic response of the TIs.

To see the actual spin distribution behind $s^z$, we have considered the on-site spin $z$ contributions and explored their relation to the on-site expectation values of $\widehat{\mathbf{L}}$ and $\widehat{\mathbf{J}}$, thereby characterizing the behavior of the valence and conduction states near the nuclei in the language of atomic orbitals. In particular, it turned out that Bi$_2$Te$_3$ is unique among the considered TIs in that the spin $s^z$ is almost exclusively determined by the on-site contribution (negative for the valence state and positive for the conduction state) coming from the atomic-like Te~$5p$ states in the outer planes of the QL---the borders of the van-der-Waals gap. We have found that the TIs respond to the field following one of two distinct scenarios, and what scenario will realize depends on whether the $s^z$ of the valence state and the $z$ projection of the total angular momentum $J^z$ on the atomic sites that have the largest weight in this state have the same (conventional response, Bi$_2$Se$_3$ and Bi$_2$Te$_2$Se) or opposite (unconventional response, Bi$_2$Te$_3$ and Sb$_2$Te$_3$) sign.

To reveal the essential physics behind different scenarios in the exchange-field effect, we have constructed a minimal four-band third-order $\mathbf{k}\cdot\mathbf{p}$ model and applied it to hypothetical  topologically trivial and non-trivial bulk insulators. Within the model, the field-induced changes are analyzed over the whole BZ, and the Chern number $\mathcal{C}(k_z)$ characterizing the topology of the modified band structure is calculated as a function of $k_z$. We have demonstrated that apart from the Weyl nodes along $k_z$ (including the double-Weyl nodes having charge $\pm2$), the conventional response may also lead to the appearance of nodes in the mirror plane away from high-symmetry lines. The unconventional response implies the creation of Weyl nodes in the mirror plane only, which results in a larger Chern number than in the conventional case.

Finally, we have shown that almost all 3D TI films up to six QLs meet the condition for the conventional response, which leads to a QAH state with $\mathcal{C}=1$. The exception are the ultra-thin films of Bi$_2$Te$_3$, where $s^z$ and $J^z$ have opposite sign, which implies the unconventional response. In particular, the two-QL film of Bi$_2$Te$_3$ undergoes a phase transition to a QAH state characterized by $\mathcal{C}=-3$. Here, like in the bulk insulators, the unconventional response is accompanied by a large Chern number. Thus, among the presented variety of the scenarios of the magnetic exchange response in the topologically trivial and non-trivial bulk crystals and thin films, we reveal the scenarios that have so far remained beyond the scope of the $\mathbf{k}\cdot\mathbf{p}$ modeling. The present \textit{ab initio} $\mathbf{k}\cdot\mathbf{p}$ approach offers a monistic treatment of a wide range of the response scenarios---a step towards deeper understanding of the electronic properties underlying magnetic phenomena in topological insulators.

\begin{acknowledgments}
We acknowledge funding from the Department of Education of the Basque Government (Grant No.~IT1164-19) and the Spanish Ministry of Science, Innovation, and Universities (Project No.~PID2019-105488GB-I00).
\end{acknowledgments}

\newpage
\pagebreak

\begin{widetext}

\appendix

\setcounter{table}{0}\renewcommand{\thetable}{\thesection~\Roman{table}}

\section{Calculation of the expectation values of $\widehat{\mathbf{S}}$, $\widehat{\mathbf{L}}$, and $\widehat{\mathbf{J}}$}\label{A_onsite}

For the valence ($n=1$) and conduction ($n=2$) Kramers-degenerate states, we represent the operator $\widehat{\bm{\mathcal{O}}}=(\widehat{\mathcal{O}}_x, \widehat{\mathcal{O}}_y, \widehat{\mathcal{O}}_z)$, where $\widehat{\bm{\mathcal{O}}} = \widehat{\mathbf{L}}$, $\widehat{\mathbf{S}}$, or $\widehat{\mathbf{J}}$, by a $2\times2$ matrix $\langle \widehat{\bm{\mathcal{O}}}\rangle_i^n$ with the elements $\langle\Psi_{n\mu}| \widehat{\bm{\mathcal{O}}} |\Psi_{n\nu}\rangle_i$ with the subscripts $\mu$ and $\nu$ numbering the members of the Kramers pairs, $\mu,\nu=\uparrow$ or $\downarrow$. The subscript $i$ of the matrix element indicates that the integration is over the MT sphere of the $i$-th atomic site of the unit cell. Due to the symmetry and the choice of the phases for the basis wave functions~\cite{Nechaev_PRBR_2016, Nechaev_PRB_2020}, at $\Gamma$, the $x$, $y$, and $z$ components of the matrices $\langle \widehat{\bm{\mathcal{O}}}\rangle_i^n$ can be expressed in terms of the Pauli matrices: $\langle \widehat{\mathcal{O}}_{x,y}\rangle_i^n =\mathcal{O}_{n,i}^{\spr} \sigma_{x,y}$ and $\langle \widehat{\mathcal{O}}_z\rangle_i^n= \mathcal{O}_{n,i}^{z} \sigma_z$.  Table~\ref{tab:table1} shows the values of $\mathcal{O}_{n,i}^{\spr}$ and $\mathcal{O}_{n,i}^{z}$, which we refer to as the on-site expectation values of the operator $\widehat{\bm{\mathcal{O}}}$, for one representative site of each type $\tau=1$, 2, and 3. In the table, we also show the weight $Q_{n,i} = \langle\Psi_{n\uparrow}| \Psi_{n\uparrow}\rangle_i$ of the $i$-th MT sphere in the state $n$ (the largest one is highlighted by italics) and the expectation value of the operator $\widehat{\mathbf{J}}^2$ calculated as $J^2_{n,i}= \langle\Psi_{n\uparrow}| \widehat{\mathbf{L}}^2 + \widehat{\mathbf{S}}^2+ 2\widehat{\mathbf{L}}\cdot \widehat{\mathbf{S}}|\Psi_{n\uparrow}\rangle_i$. The on-site expectation values listed in Table~\ref{tab:table1} are visualized in Fig.~\ref{LSJ_on_site} by the vectors $\bm{\mathcal{O}}_{n,i} = (\mathcal{O}_{n,i}^{\spr}, \mathcal{O}_{n,i}^{z})/Q_{n,i}$ on the two-dimensional ($\spr,\,z$) plane.

\begin{table*}[t]
\caption{\label{tab:table1} On-site expectation values of the orbital angular momentum $\widehat{\mathbf{L}}$, spin $\widehat{\mathbf{S}}$, and total angular momentum $\widehat{\mathbf{J}}$ in Rydberg atomic units for the bulk crystals of Bi$_2$Se$_3$, Bi$_2$Te$_2$Se, Bi$_2$Te$_3$, and Sb$_2$Te$_3$.}
\begin{ruledtabular}
\begin{tabular}{crrrrrrrrrrrrrrrr}
$i$ & \multicolumn{1}{c}{$L_{1,i}^{\spr}$} & \multicolumn{1}{c}{$L_{1,i}^z$} & \multicolumn{1}{c}{$S_{1,i}^{\spr}$} &
          \multicolumn{1}{c}{$S_{1,i}^z$}          & \multicolumn{1}{c}{$J_{1,i}^{\spr}$} & \multicolumn{1}{c}{$J_{1,i}^z$} &
          \multicolumn{1}{c}{$J_{1,i}^2$}           & \multicolumn{1}{c}{$Q_{1,i}$}  & \multicolumn{1}{c}{$L_{2,i}^{\spr}$} & \multicolumn{1}{c}{$L_{2,i}^z$}          & \multicolumn{1}{c}{$S_{2,i}^{\spr}$} & \multicolumn{1}{c}{$S_{2,i}^z$}  & \multicolumn{1}{c}{$J_{2,i}^{\spr}$}  & \multicolumn{1}{c}{$J_{2,i}^z$}            & \multicolumn{1}{c}{$J_{2,i}^2$} & \multicolumn{1}{c}{$Q_{2,i}$}\\
\hline
&\multicolumn{16}{c}{\bf Bi$_2$Se$_3$}  \\
Bi         & 0.111   & 0.047               &-0.065    & 0.038              & 0.045    &0.085         & 0.274 & \multicolumn{1}{c}{\it0.185}
            & 0.002   & 0.000	               &-0.030	 & 0.030	              &-0.029	&0.030          & 0.122 & 0.061 \\
Se$_1$ & 0.017  & 0.073                &-0.014   & -0.026              & 0.003    &0.046         & 0.213 & 0.111
            &-0.059  &0.006                 &-0.157	& 0.154	              &-0.216	&0.160          &1.006 &\multicolumn{1}{c}{\it0.320} \\
Se$_2$ &-0.023 &-0.007                &-0.033   &0.027                 &-0.056   &0.020          &0.285 &0.077
            &-0.012 &0.001	               &-0.029	&0.028	              &-0.041	&0.029           &0.234 & 0.059 \\
&\multicolumn{16}{c}{\bf Bi$_2$Te$_2$Se} \\
Bi        & 0.117  &0.064	               &-0.057	& 0.020	           & 0.061	&0.084              &0.257 & \multicolumn{1}{c}{\it0.186}
            &-0.006 &0.003	              &-0.035	&0.032              &-0.041	&0.034	            &0.189 &0.075 \\
Te$_1$ & 0.042  &0.091	             &-0.018	&-0.033	               &0.023	&0.058               &0.235 &0.140
            &-0.106 &0.021	             &-0.131	&0.119	               &-0.237	&0.141	            &0.969 &\multicolumn{1}{c}{\it0.284}\\
Se$_2$ &-0.018 &-0.008	             & -0.020 & 0.013	               &-0.037	&0.006               &0.216 &0.052
            &-0.016 &0.003	             &-0.025	&0.024	               &-0.041	&0.027	            &0.241 &0.054\\
&\multicolumn{16}{c}{\bf Bi$_2$Te$_3$} \\
Bi         &0.131  &0.085	              &-0.049	 &0.001	                &0.081	&0.085           &0.239 &\multicolumn{1}{c}{\it0.196}
             &-0.002 &0.001	              &-0.032	 &0.029	                &-0.034	&0.030	         &0.175 &0.069 \\
Te$_1$ & 0.022  &0.109	             &-0.009	     &-0.053	                &0.012	&0.057            &0.269 &0.142
             &-0.104 &0.022	             &-0.125	      &0.114	                &-0.229	&0.136	        &0.935 &\multicolumn{1}{c}{\it0.274} \\
Te$_2$ & -0.011 &-0.006	         &-0.018	      &0.013	                &-0.029	&0.007                &0.150  &0.046
             &-0.027 &0.009	             &-0.022	       &0.018	            &-0.050	&0.027	              &0.231 &0.054\\
&\multicolumn{16}{c}{\bf Sb$_2$Te$_3$} \\
Sb         &0.146  &0.099	             &-0.055	       &0.001	            &0.091	&0.100            &0.241 &\multicolumn{1}{c}{\it0.217}
             &0.001  &0.003	             &-0.041	        &0.038	            &-0.040	&0.041	          &0.179 &0.088\\
Te$_1$ & 0.019  &0.050	             &-0.015	        &-0.020	            &0.004	&0.031             &0.227 &0.098
             &-0.056 &0.006	             &-0.119	        &0.116	            &-0.175	&0.122	         &0.812 &\multicolumn{1}{c}{\it0.245}\\
Te$_2$ &-0.016 &-0.010	             &-0.023	        &0.015	            &-0.039	&0.005            &0.225 &0.061
             &-0.042 &0.015	             &-0.033	        &0.026	            &-0.075	&0.040	          &0.343 &0.081
\end{tabular}
\end{ruledtabular}
\end{table*}

\newpage
\pagebreak

\section{Parameters of the four-band Hamiltonian}\label{B_parameters}

The parameters in Eq.~(\ref{kpHamForBulk}) for the bulk crystals are listed in Table~\ref{tab:table2}. For all the films considered, the parameters in Eqs.~(\ref{HamTwoSurfaces}) and (\ref{SpinTwoSurfaces}) are presented in Tables~\ref{tab:table3}-\ref{tab:table5}. Note that some values slightly differ from those in Ref.~\cite{Nechaev_PRBR_2016} because they are derived from a numerically different all-electron band structure. Regarding the one-QL films, we would like to note that only in Bi$_2$Te$_3$ the chosen basis state $|\Psi_{1}\rangle$ is not the highest valence state but the second highest one, see Ref.~\cite{Nechaev_PRBR_2016}. Both these states are largely localized on the Te$_1$ sites, where, as suggested by the behavior of the vectors $\mathbf{L}_{1,\rtm{Te}_1}$, $\mathbf{S}_{1,\rtm{Te}_1}$, and $\mathbf{J}_{1,\rtm{Te}_1 }$, $|\Psi_{1}\rangle$ is a purely Te-atomic $\left|\frac{3}{2},\pm\frac{1}{2}\right\rangle$ state, while the highest valence state exhibits all the characteristics of an atomic $\left|\frac{3}{2},\pm\frac{3}{2}\right\rangle$ state.  At the same time, $|\Psi_{2}\rangle$ is the lowest conduction state, as in all our calculations. Here, this state is almost equally localized on Bi and Te$_1$ sites (the difference between $Q_{2,\rtm{Bi}}$ and $Q_{2,\rtm{Te}_1}$ is less than 1\%), and its vectors $\mathbf{L}_{2,i}$, $\mathbf{S}_{2,i}$, and $\mathbf{J}_{2,i}$ are similar to those of the state $n=1$ in the bulk TIs.

\begin{table*}[tbp]
\caption{\label{tab:table2} Parameters of the four-band $\mathbf{k}\cdot\mathbf{p}$ Hamiltonian~(\ref{kpHamForBulk}) based on calculations for bulk crystals of Bi$_2$Se$_3$, Bi$_2$Te$_2$Se, Bi$_2$Te$_3$, and Sb$_2$Te$_3$ with the lattice parameters $a=18.5968$, 19.3792, 19.7911, and 19.7023~a.u., respectively. We use Rydberg atomic units: $\hbar=2m_0=e^2/2=1$.}
\begin{ruledtabular}
\begin{tabular}{ldddd}
&\multicolumn{1}{c}{Bi$_2$Se$_3$}        & \multicolumn{1}{c}{Bi$_2$Te$_2$Se}       & \multicolumn{1}{c}{Bi$_2$Te$_3$}          & \multicolumn{1}{c}{Sb$_2$Te$_3$} \\
  \hline
$\epsilon_1$                          &    -0.010                     &   -0.030                        &    -0.034                 & -0.014                 \\
$\epsilon_2$                          &      0.016                     &     0.020                        &     0.011                 &   0.013                 \\
$\alpha^{(1)}$                      &   -0.348                       &    -0.516                       &    -0.554                 & -0.514                 \\
$\beta^{(1)}$                         &   0.255                         &     0.233                        &    0.125                  & 0.163                   \\
$\alpha^{(3)}$                       &      -2.22                       &    144.84                      &    533.02                &  46.17                  \\
$\beta^{(3)}$                         &   -1.54                           &    -0.98                         &    1.94                     & -2.33                    \\
$\tilde{\alpha}^{(3)}$          &    19.61                         &    40.82                        &    41.01                  &  105.19                \\
$\tilde{\beta}^{(3)}$            &   -49.31                        &   -149.82                     &    -236.88              & -147.26                \\
$\theta$                                   &    6.16                          &     41.92                         &  75.75                    & 29.91                 \\
$\eta$                                       &   -22.48                         &    -61.78                       &   -95.37                & -70.39                  \\
$\delta$                                    &   36.77                          &   180.45                     &  427.12                   & 214.70                 \\
$M_1^{\spr}$                          &  11.50                            &   44.08                       &   109.51                  &  11.64                  \\
$M_2^{\spr}$                          &  -4.08                            &    -6.53                       &   -5.94                     &  -14.95                 \\
$M_1^z$                                   &  1.27                              &     2.74                        &   3.13                      &  1.45                     \\
$M_2^z$                                   &   -0.52                            &    -2.33                      &   -1.75                      &  -8.89                   \\
$s^{\spr}_1 $                           &   -0.63                           &   -0.52                        &  -0.39                      &  -0.49                    \\
$s^{\spr}_2$                            &   -0.98                           &   -0.92                        &   -0.91                     & -0.94                     \\
$ s^{z}_1$                                 &   0.27                             &    0.04                        &  -0.22                       &  -0.02                   \\
$s^{z}_2$                                   &   0.96                             &    0.84                        &   0.82                        &   0.89
\end{tabular}
\end{ruledtabular}
\end{table*}

\begin{table*}[tbp]
\caption{\label{tab:table3} Parameters of the Hamiltonian~(\ref{HamTwoSurfaces}) for Bi$_2$Se$_3$ slabs with the lattice parameter $a=7.8187$~a.u. in Rydberg atomic units (except $\epsilon$ and $\Delta \epsilon$ given in eV).}
\begin{ruledtabular}
\begin{tabular}{ldddddd}
               & \multicolumn{1}{c}{1QL}
               & \multicolumn{1}{c}{2QL}
               & \multicolumn{1}{c}{3QL}
               & \multicolumn{1}{c}{4QL}
               & \multicolumn{1}{c}{5QL}
               & \multicolumn{1}{c}{6QL}       \\
\hline
$\epsilon$                          &   -0.027  &  -0.053          &  -0.103      &  -0.111     &  -0.112      &  -0.117      \\
$\alpha^{(1)}$                  &  -0.264   &   -0.167    &  -0.175   &  -0.179  & -0.177   &  -0.174   \\
$\alpha^{(3)}$                  &  26.74    &  -23.88      & -29.31    &  -31.54 & -29.32    &  -27.81   \\
$\theta$                             &  11.57    &    3.69        &   4.94      &    5.28    &   5.67      &   5.80     \\
$M^{\spr}$                        &  9.54      &    7.91        &   7.91      &    7.82    &   7.90      &   7.96     \\
$\bar{s}^{\spr}$                          &  0.79      &    0.73          &   0.71      &    0.71    &   0.70      &  0.70     \\
$ \bar{s}^{z}$                              &   0.58     &     0.46         &   0.42       &   0.42     &   0.40      &  0.40     \\
\hline
$\Delta \epsilon$             &  -0.376  &   -0.050          & -0.014         &    -0.004         &  -0.001           &    0.000          \\
$\Delta M^{\spr}$           &   1.98     &    -0.70        &  -0.40     &    -0.17   &  -0.02      &  0.02     \\
$\eta$                                &  27.00    &    -6.87        &  -6.82      &   -4.13    &  -2.14      &  -0.93     \\
$\Delta s^{\spr}$            &   0.12      &    0.08         &  0.05        &   0.03     &   0.01      &  0.01     \\
$\Delta s^{z}$                  &   0.24      &     0.16         &  0.10        &   0.05     &   0.02      &  0.01
\end{tabular}
\end{ruledtabular}
\end{table*}

\begin{table*}[tbp]
\caption{\label{tab:table4} Same as in Table~\ref{tab:table3}, but for Bi$_2$Te$_3$ with the lattice parameter $a=8.2870$~a.u.}
\begin{ruledtabular}
\begin{tabular}{ldddddd}
               & \multicolumn{1}{c}{1QL}
               & \multicolumn{1}{c}{2QL}
               & \multicolumn{1}{c}{3QL}
               & \multicolumn{1}{c}{4QL}
               & \multicolumn{1}{c}{5QL}
               & \multicolumn{1}{c}{6QL}       \\
\hline
$\epsilon$                  &   -0.110      &  -0.109        &  -0.170      &  -0.186     &  -0.195      &  -0.199      \\
$\alpha^{(1)}$          &  -0.400       &   -0.151    &  -0.143   &  -0.137  & -0.147    &  -0.146   \\
$\alpha^{(3)}$          &   -25.98      &   42.95      &  -0.07     &     -8.76  &   -0.89    &   -4.86    \\
$\theta$                      &   30.31       &   50.04      &   25.39    &   19.41   &  23.58    &  22.84     \\
$M^{\spr}$                &   15.00        &   16.58      &  12.34     &   12.23    &  12.19   & 12.23     \\
$\bar{s}^{\spr}$                  &  -0.52         &    0.52         &   0.58      &    0.61    &   0.60      &  0.60     \\
$ \bar{s}^{z}$                       &   0.05          &     0.04         &   0.16       &   0.22     &   0.19      &  0.20     \\
\hline
$\Delta \epsilon$      &  -0.283      &   -0.057          &     -0.004         &    -0.007         &  -0.004          &    0.000          \\
$\Delta M^{\spr}$    &   3.90         &    4.81         &   0.27      &     0.11   &   0.20      &  -0.02     \\
$\eta$                          &   8.23         &   14.84       &   5.17      &    4.33    &   4.89     &  -0.37     \\
$\Delta s^{\spr}$      &  -0.06        &   -0.10         & -0.09       &   0.04     &  -0.02     &  0.02     \\
$\Delta s^{z}$           &   0.12         &    -0.21         & -0.19        &  0.09     &  -0.03      &  0.04
\end{tabular}
\end{ruledtabular}
\end{table*}

\begin{table*}
\caption{\label{tab:table5} Same as in Table~\ref{tab:table3}, but for Sb$_2$Te$_3$ with the lattice parameter  $a=8.0312$~a.u.}
\begin{ruledtabular}
\begin{tabular}{ldddddd}
               & \multicolumn{1}{c}{1QL}
               & \multicolumn{1}{c}{2QL}
               & \multicolumn{1}{c}{3QL}
               & \multicolumn{1}{c}{4QL}
               & \multicolumn{1}{c}{5QL}
               & \multicolumn{1}{c}{6QL}       \\
\hline
$\epsilon$                  &  -0.031      &  -0.061          &  -0.051     &   -0.087       &    -0.093      &  -0.101      \\
$\alpha^{(1)}$          &  -0.379      &   -0.290    &  -0.272   &  -0.258  & -0.271   &  -0.267   \\
$\alpha^{(3)}$          &  110.25     &    -5.42      & -15.40   &  -21.14  & -33.33    &  -32.14   \\
$\theta$                     &    34.04      &   20.28       &  40.53  &  52.30    &  52.71    &  52.05     \\
$M^{\spr}$                &    9.52        &    8.61        &   7.21    &    7.28    &   7.37      &   7.45     \\
$\bar{s}^{\spr}$                  &    0.73        &    0.73          &   0.66   &    0.62    &   0.62      &  0.63     \\
$ \bar{s}^{z}$                       &    0.47        &     0.46         &   0.32    &   0.24     &   0.25      &  0.26     \\
\hline
$\Delta \epsilon$     &    -0.349     &   -0.056          &   -0.002          &   -0.010       &  -0.007          &   -0.002          \\
$\Delta M^{\spr}$   &      1.74       &     1.38        &   -2.44    &    -2.50   &  -0.80      &  0.63     \\
$\eta$                        &     41.07      &   -20.87      & 30.65    & 19.28   &  1.71      &  -5.74     \\
$\Delta s^{\spr}$    &     0.12         &    0.19         &  -0.16     &  -0.08     &   0.00      &  0.03     \\
$\Delta s^{z}$          &     0.23         &     0.37         &  -0.31      &  -0.17     &  -0.01      &  0.06
\end{tabular}
\end{ruledtabular}
\end{table*}

\end{widetext}

\newpage
\pagebreak


\begin{thebibliography}{46}%
\makeatletter
\providecommand \@ifxundefined [1]{%
 \@ifx{#1\undefined}
}%
\providecommand \@ifnum [1]{%
 \ifnum #1\expandafter \@firstoftwo
 \else \expandafter \@secondoftwo
 \fi
}%
\providecommand \@ifx [1]{%
 \ifx #1\expandafter \@firstoftwo
 \else \expandafter \@secondoftwo
 \fi
}%
\providecommand \natexlab [1]{#1}%
\providecommand \enquote  [1]{``#1''}%
\providecommand \bibnamefont  [1]{#1}%
\providecommand \bibfnamefont [1]{#1}%
\providecommand \citenamefont [1]{#1}%
\providecommand \href@noop [0]{\@secondoftwo}%
\providecommand \href [0]{\begingroup \@sanitize@url \@href}%
\providecommand \@href[1]{\@@startlink{#1}\@@href}%
\providecommand \@@href[1]{\endgroup#1\@@endlink}%
\providecommand \@sanitize@url [0]{\catcode `\\12\catcode `\$12\catcode
  `\&12\catcode `\#12\catcode `\^12\catcode `\_12\catcode `\%12\relax}%
\providecommand \@@startlink[1]{}%
\providecommand \@@endlink[0]{}%
\providecommand \url  [0]{\begingroup\@sanitize@url \@url }%
\providecommand \@url [1]{\endgroup\@href {#1}{\urlprefix }}%
\providecommand \urlprefix  [0]{URL }%
\providecommand \Eprint [0]{\href }%
\providecommand \doibase [0]{https://doi.org/}%
\providecommand \selectlanguage [0]{\@gobble}%
\providecommand \bibinfo  [0]{\@secondoftwo}%
\providecommand \bibfield  [0]{\@secondoftwo}%
\providecommand \translation [1]{[#1]}%
\providecommand \BibitemOpen [0]{}%
\providecommand \bibitemStop [0]{}%
\providecommand \bibitemNoStop [0]{.\EOS\space}%
\providecommand \EOS [0]{\spacefactor3000\relax}%
\providecommand \BibitemShut  [1]{\csname bibitem#1\endcsname}%
\let\auto@bib@innerbib\@empty
\bibitem [{\citenamefont {Weng}\ \emph {et~al.}(2015)\citenamefont {Weng},
  \citenamefont {Yu}, \citenamefont {Hu}, \citenamefont {Dai},\ and\
  \citenamefont {Fang}}]{Weng_AP_2015}%
  \BibitemOpen
  \bibfield  {author} {\bibinfo {author} {\bibfnamefont {H.}~\bibnamefont
  {Weng}}, \bibinfo {author} {\bibfnamefont {R.}~\bibnamefont {Yu}}, \bibinfo
  {author} {\bibfnamefont {X.}~\bibnamefont {Hu}}, \bibinfo {author}
  {\bibfnamefont {X.}~\bibnamefont {Dai}},\ and\ \bibinfo {author}
  {\bibfnamefont {Z.}~\bibnamefont {Fang}},\ }\bibfield  {title} {\bibinfo
  {title} {{Quantum anomalous Hall effect and related topological electronic
  states}},\ }\href {https://doi.org/10.1080/00018732.2015.1068524} {\bibfield
  {journal} {\bibinfo  {journal} {Advances in Physics}\ }\textbf {\bibinfo
  {volume} {64}},\ \bibinfo {pages} {227} (\bibinfo {year} {2015})}\BibitemShut
  {NoStop}%
\bibitem [{\citenamefont {Liu}\ \emph {et~al.}(2016)\citenamefont {Liu},
  \citenamefont {Zhang},\ and\ \citenamefont {Qi}}]{Liu_ARCMP_2016}%
  \BibitemOpen
  \bibfield  {author} {\bibinfo {author} {\bibfnamefont {C.-X.}\ \bibnamefont
  {Liu}}, \bibinfo {author} {\bibfnamefont {S.-C.}\ \bibnamefont {Zhang}},\
  and\ \bibinfo {author} {\bibfnamefont {X.-L.}\ \bibnamefont {Qi}},\
  }\bibfield  {title} {\bibinfo {title} {{The Quantum Anomalous Hall Effect:
  Theory and Experiment}},\ }\href
  {https://doi.org/10.1146/annurev-conmatphys-031115-011417} {\bibfield
  {journal} {\bibinfo  {journal} {Annual Review of Condensed Matter Physics}\
  }\textbf {\bibinfo {volume} {7}},\ \bibinfo {pages} {301} (\bibinfo {year}
  {2016})}\BibitemShut {NoStop}%
\bibitem [{\citenamefont {Tokura}\ \emph {et~al.}(2019)\citenamefont {Tokura},
  \citenamefont {Yasuda},\ and\ \citenamefont {Tsukazaki}}]{Tokura_NRP_2019}%
  \BibitemOpen
  \bibfield  {author} {\bibinfo {author} {\bibfnamefont {Y.}~\bibnamefont
  {Tokura}}, \bibinfo {author} {\bibfnamefont {K.}~\bibnamefont {Yasuda}},\
  and\ \bibinfo {author} {\bibfnamefont {A.}~\bibnamefont {Tsukazaki}},\
  }\bibfield  {title} {\bibinfo {title} {Magnetic topological insulators},\
  }\href {https://doi.org/10.1038/s42254-018-0011-5} {\bibfield  {journal}
  {\bibinfo  {journal} {Nature Reviews Physics}\ }\textbf {\bibinfo {volume}
  {1}},\ \bibinfo {pages} {126} (\bibinfo {year} {2019})}\BibitemShut {NoStop}%
\bibitem [{\citenamefont {Kane}\ and\ \citenamefont
  {Mele}(2005{\natexlab{a}})}]{Kane_PRL_2005_1}%
  \BibitemOpen
  \bibfield  {author} {\bibinfo {author} {\bibfnamefont {C.~L.}\ \bibnamefont
  {Kane}}\ and\ \bibinfo {author} {\bibfnamefont {E.~J.}\ \bibnamefont
  {Mele}},\ }\bibfield  {title} {\bibinfo {title} {{Quantum Spin Hall Effect in
  Graphene}},\ }\href {https://doi.org/10.1103/PhysRevLett.95.226801}
  {\bibfield  {journal} {\bibinfo  {journal} {Phys. Rev. Lett.}\ }\textbf
  {\bibinfo {volume} {95}},\ \bibinfo {pages} {226801} (\bibinfo {year}
  {2005}{\natexlab{a}})}\BibitemShut {NoStop}%
\bibitem [{\citenamefont {Kane}\ and\ \citenamefont
  {Mele}(2005{\natexlab{b}})}]{Kane_PRL_2005_2}%
  \BibitemOpen
  \bibfield  {author} {\bibinfo {author} {\bibfnamefont {C.~L.}\ \bibnamefont
  {Kane}}\ and\ \bibinfo {author} {\bibfnamefont {E.~J.}\ \bibnamefont
  {Mele}},\ }\bibfield  {title} {\bibinfo {title} {{${Z}_{2}$ Topological Order
  and the Quantum Spin Hall Effect}},\ }\href
  {https://doi.org/10.1103/PhysRevLett.95.146802} {\bibfield  {journal}
  {\bibinfo  {journal} {Phys. Rev. Lett.}\ }\textbf {\bibinfo {volume} {95}},\
  \bibinfo {pages} {146802} (\bibinfo {year} {2005}{\natexlab{b}})}\BibitemShut
  {NoStop}%
\bibitem [{\citenamefont {Bernevig}\ \emph {et~al.}(2006)\citenamefont
  {Bernevig}, \citenamefont {Hughes},\ and\ \citenamefont
  {Zhang}}]{Bernevig_Science_2006}%
  \BibitemOpen
  \bibfield  {author} {\bibinfo {author} {\bibfnamefont {B.~A.}\ \bibnamefont
  {Bernevig}}, \bibinfo {author} {\bibfnamefont {T.~L.}\ \bibnamefont
  {Hughes}},\ and\ \bibinfo {author} {\bibfnamefont {S.-C.}\ \bibnamefont
  {Zhang}},\ }\bibfield  {title} {\bibinfo {title} {{Quantum Spin Hall Effect
  and Topological Phase Transition in HgTe Quantum Wells}},\ }\href
  {https://doi.org/10.1126/science.1133734} {\bibfield  {journal} {\bibinfo
  {journal} {Science}\ }\textbf {\bibinfo {volume} {314}},\ \bibinfo {pages}
  {1757} (\bibinfo {year} {2006})}\BibitemShut {NoStop}%
\bibitem [{\citenamefont {Haldane}(1988)}]{Haldane_PRL_1988}%
  \BibitemOpen
  \bibfield  {author} {\bibinfo {author} {\bibfnamefont {F.~D.~M.}\
  \bibnamefont {Haldane}},\ }\bibfield  {title} {\bibinfo {title} {{Model for a
  Quantum Hall Effect without Landau Levels: Condensed-Matter Realization of
  the "Parity Anomaly"}},\ }\href {https://doi.org/10.1103/PhysRevLett.61.2015}
  {\bibfield  {journal} {\bibinfo  {journal} {Phys. Rev. Lett.}\ }\textbf
  {\bibinfo {volume} {61}},\ \bibinfo {pages} {2015} (\bibinfo {year}
  {1988})}\BibitemShut {NoStop}%
\bibitem [{\citenamefont {Onoda}\ and\ \citenamefont
  {Nagaosa}(2003)}]{Onoda_PRL_2003}%
  \BibitemOpen
  \bibfield  {author} {\bibinfo {author} {\bibfnamefont {M.}~\bibnamefont
  {Onoda}}\ and\ \bibinfo {author} {\bibfnamefont {N.}~\bibnamefont
  {Nagaosa}},\ }\bibfield  {title} {\bibinfo {title} {{Quantized Anomalous Hall
  Effect in Two-Dimensional Ferromagnets: Quantum Hall Effect in Metals}},\
  }\href {https://doi.org/10.1103/PhysRevLett.90.206601} {\bibfield  {journal}
  {\bibinfo  {journal} {Phys. Rev. Lett.}\ }\textbf {\bibinfo {volume} {90}},\
  \bibinfo {pages} {206601} (\bibinfo {year} {2003})}\BibitemShut {NoStop}%
\bibitem [{\citenamefont {Liu}\ \emph {et~al.}(2008)\citenamefont {Liu},
  \citenamefont {Qi}, \citenamefont {Dai}, \citenamefont {Fang},\ and\
  \citenamefont {Zhang}}]{Liu_PRL_2008}%
  \BibitemOpen
  \bibfield  {author} {\bibinfo {author} {\bibfnamefont {C.-X.}\ \bibnamefont
  {Liu}}, \bibinfo {author} {\bibfnamefont {X.-L.}\ \bibnamefont {Qi}},
  \bibinfo {author} {\bibfnamefont {X.}~\bibnamefont {Dai}}, \bibinfo {author}
  {\bibfnamefont {Z.}~\bibnamefont {Fang}},\ and\ \bibinfo {author}
  {\bibfnamefont {S.-C.}\ \bibnamefont {Zhang}},\ }\bibfield  {title} {\bibinfo
  {title} {{Quantum Anomalous Hall Effect in
  ${\mathrm{Hg}}_{1\ensuremath{-}y}{\mathrm{Mn}}_{y}\mathrm{Te}$ Quantum
  Wells}},\ }\href {https://doi.org/10.1103/PhysRevLett.101.146802} {\bibfield
  {journal} {\bibinfo  {journal} {Phys. Rev. Lett.}\ }\textbf {\bibinfo
  {volume} {101}},\ \bibinfo {pages} {146802} (\bibinfo {year}
  {2008})}\BibitemShut {NoStop}%
\bibitem [{\citenamefont {Yu}\ \emph {et~al.}(2010)\citenamefont {Yu},
  \citenamefont {Zhang}, \citenamefont {Zhang}, \citenamefont {Zhang},
  \citenamefont {Dai},\ and\ \citenamefont {Fang}}]{Yu_Science_2010}%
  \BibitemOpen
  \bibfield  {author} {\bibinfo {author} {\bibfnamefont {R.}~\bibnamefont
  {Yu}}, \bibinfo {author} {\bibfnamefont {W.}~\bibnamefont {Zhang}}, \bibinfo
  {author} {\bibfnamefont {H.-J.}\ \bibnamefont {Zhang}}, \bibinfo {author}
  {\bibfnamefont {S.-C.}\ \bibnamefont {Zhang}}, \bibinfo {author}
  {\bibfnamefont {X.}~\bibnamefont {Dai}},\ and\ \bibinfo {author}
  {\bibfnamefont {Z.}~\bibnamefont {Fang}},\ }\bibfield  {title} {\bibinfo
  {title} {{Quantized Anomalous Hall Effect in Magnetic Topological
  Insulators}},\ }\href {https://doi.org/10.1126/science.1187485} {\bibfield
  {journal} {\bibinfo  {journal} {Science}\ }\textbf {\bibinfo {volume}
  {329}},\ \bibinfo {pages} {61} (\bibinfo {year} {2010})}\BibitemShut
  {NoStop}%
\bibitem [{\citenamefont {Lu}\ \emph {et~al.}(2013)\citenamefont {Lu},
  \citenamefont {Zhao},\ and\ \citenamefont {Shen}}]{Lu_PRL_2013}%
  \BibitemOpen
  \bibfield  {author} {\bibinfo {author} {\bibfnamefont {H.-Z.}\ \bibnamefont
  {Lu}}, \bibinfo {author} {\bibfnamefont {A.}~\bibnamefont {Zhao}},\ and\
  \bibinfo {author} {\bibfnamefont {S.-Q.}\ \bibnamefont {Shen}},\ }\bibfield
  {title} {\bibinfo {title} {{Quantum Transport in Magnetic Topological
  Insulator Thin Films}},\ }\href
  {https://doi.org/10.1103/PhysRevLett.111.146802} {\bibfield  {journal}
  {\bibinfo  {journal} {Phys. Rev. Lett.}\ }\textbf {\bibinfo {volume} {111}},\
  \bibinfo {pages} {146802} (\bibinfo {year} {2013})}\BibitemShut {NoStop}%
\bibitem [{\citenamefont {Wang}\ \emph {et~al.}(2015)\citenamefont {Wang},
  \citenamefont {Lian},\ and\ \citenamefont {Zhang}}]{Wang_PRL_2015}%
  \BibitemOpen
  \bibfield  {author} {\bibinfo {author} {\bibfnamefont {J.}~\bibnamefont
  {Wang}}, \bibinfo {author} {\bibfnamefont {B.}~\bibnamefont {Lian}},\ and\
  \bibinfo {author} {\bibfnamefont {S.-C.}\ \bibnamefont {Zhang}},\ }\bibfield
  {title} {\bibinfo {title} {{Electrically Tunable Magnetism in Magnetic
  Topological Insulators}},\ }\href
  {https://doi.org/10.1103/PhysRevLett.115.036805} {\bibfield  {journal}
  {\bibinfo  {journal} {Phys. Rev. Lett.}\ }\textbf {\bibinfo {volume} {115}},\
  \bibinfo {pages} {036805} (\bibinfo {year} {2015})}\BibitemShut {NoStop}%
\bibitem [{\citenamefont {Zhang}\ \emph {et~al.}(2020)\citenamefont {Zhang},
  \citenamefont {Wu},\ and\ \citenamefont {Das~Sarma}}]{Zhang_PRL_2020}%
  \BibitemOpen
  \bibfield  {author} {\bibinfo {author} {\bibfnamefont {R.-X.}\ \bibnamefont
  {Zhang}}, \bibinfo {author} {\bibfnamefont {F.}~\bibnamefont {Wu}},\ and\
  \bibinfo {author} {\bibfnamefont {S.}~\bibnamefont {Das~Sarma}},\ }\bibfield
  {title} {\bibinfo {title} {{M\"obius Insulator and Higher-Order Topology in
  ${\mathrm{MnBi}}_{2n}{\mathrm{Te}}_{3n+1}$}},\ }\href
  {https://doi.org/10.1103/PhysRevLett.124.136407} {\bibfield  {journal}
  {\bibinfo  {journal} {Phys. Rev. Lett.}\ }\textbf {\bibinfo {volume} {124}},\
  \bibinfo {pages} {136407} (\bibinfo {year} {2020})}\BibitemShut {NoStop}%
\bibitem [{\citenamefont {Zhang}\ \emph {et~al.}(2013)\citenamefont {Zhang},
  \citenamefont {Chang}, \citenamefont {Tang}, \citenamefont {Zhang},
  \citenamefont {Feng}, \citenamefont {Li}, \citenamefont {Wang}, \citenamefont
  {Chen}, \citenamefont {Liu}, \citenamefont {Duan}, \citenamefont {He},
  \citenamefont {Xue}, \citenamefont {Ma},\ and\ \citenamefont
  {Wang}}]{Zhang_Science_2013}%
  \BibitemOpen
  \bibfield  {author} {\bibinfo {author} {\bibfnamefont {J.}~\bibnamefont
  {Zhang}}, \bibinfo {author} {\bibfnamefont {C.-Z.}\ \bibnamefont {Chang}},
  \bibinfo {author} {\bibfnamefont {P.}~\bibnamefont {Tang}}, \bibinfo {author}
  {\bibfnamefont {Z.}~\bibnamefont {Zhang}}, \bibinfo {author} {\bibfnamefont
  {X.}~\bibnamefont {Feng}}, \bibinfo {author} {\bibfnamefont {K.}~\bibnamefont
  {Li}}, \bibinfo {author} {\bibfnamefont {L.-l.}\ \bibnamefont {Wang}},
  \bibinfo {author} {\bibfnamefont {X.}~\bibnamefont {Chen}}, \bibinfo {author}
  {\bibfnamefont {C.}~\bibnamefont {Liu}}, \bibinfo {author} {\bibfnamefont
  {W.}~\bibnamefont {Duan}}, \bibinfo {author} {\bibfnamefont {K.}~\bibnamefont
  {He}}, \bibinfo {author} {\bibfnamefont {Q.-K.}\ \bibnamefont {Xue}},
  \bibinfo {author} {\bibfnamefont {X.}~\bibnamefont {Ma}},\ and\ \bibinfo
  {author} {\bibfnamefont {Y.}~\bibnamefont {Wang}},\ }\bibfield  {title}
  {\bibinfo {title} {{Topology-Driven Magnetic Quantum Phase Transition in
  Topological Insulators}},\ }\href {https://doi.org/10.1126/science.1230905}
  {\bibfield  {journal} {\bibinfo  {journal} {Science}\ }\textbf {\bibinfo
  {volume} {339}},\ \bibinfo {pages} {1582} (\bibinfo {year}
  {2013})}\BibitemShut {NoStop}%
\bibitem [{\citenamefont {Zhang}\ \emph {et~al.}(2009)\citenamefont {Zhang},
  \citenamefont {Liu}, \citenamefont {Qi}, \citenamefont {Dai}, \citenamefont
  {Fang},\ and\ \citenamefont {Zhang}}]{Zhang_NATPHYS_2009}%
  \BibitemOpen
  \bibfield  {author} {\bibinfo {author} {\bibfnamefont {H.}~\bibnamefont
  {Zhang}}, \bibinfo {author} {\bibfnamefont {C.-X.}\ \bibnamefont {Liu}},
  \bibinfo {author} {\bibfnamefont {X.-L.}\ \bibnamefont {Qi}}, \bibinfo
  {author} {\bibfnamefont {X.}~\bibnamefont {Dai}}, \bibinfo {author}
  {\bibfnamefont {Z.}~\bibnamefont {Fang}},\ and\ \bibinfo {author}
  {\bibfnamefont {S.-C.}\ \bibnamefont {Zhang}},\ }\bibfield  {title} {\bibinfo
  {title} {{Topological insulators in Bi$_2$Se$_3$, Bi$_2$Te$_3$ and
  Sb$_2$Te$_3$ with a single Dirac cone on the surface}},\ }\href
  {https://doi.org/10.1038/nphys1270} {\bibfield  {journal} {\bibinfo
  {journal} {Nature Physics}\ }\textbf {\bibinfo {volume} {5}},\ \bibinfo
  {pages} {438} (\bibinfo {year} {2009})}\BibitemShut {NoStop}%
\bibitem [{\citenamefont {Nechaev}\ and\ \citenamefont
  {Krasovskii}(2016)}]{Nechaev_PRBR_2016}%
  \BibitemOpen
  \bibfield  {author} {\bibinfo {author} {\bibfnamefont {I.~A.}\ \bibnamefont
  {Nechaev}}\ and\ \bibinfo {author} {\bibfnamefont {E.~E.}\ \bibnamefont
  {Krasovskii}},\ }\bibfield  {title} {\bibinfo {title} {{Relativistic
  $\mathrm{k}\ifmmode\cdot\else\textperiodcentered\fi{}\mathrm{p}$ Hamiltonians
  for centrosymmetric topological insulators from \textit{ab initio} wave
  functions}},\ }\href {https://doi.org/10.1103/PhysRevB.94.201410} {\bibfield
  {journal} {\bibinfo  {journal} {Phys. Rev. B}\ }\textbf {\bibinfo {volume}
  {94}},\ \bibinfo {pages} {201410(R)} (\bibinfo {year} {2016})}\BibitemShut
  {NoStop}%
\bibitem [{\citenamefont {Nechaev}\ \emph {et~al.}(2017)\citenamefont
  {Nechaev}, \citenamefont {Eremeev}, \citenamefont {Krasovskii}, \citenamefont
  {Echenique},\ and\ \citenamefont {Chulkov}}]{Nechaev_SciRep_2017}%
  \BibitemOpen
  \bibfield  {author} {\bibinfo {author} {\bibfnamefont {I.~A.}\ \bibnamefont
  {Nechaev}}, \bibinfo {author} {\bibfnamefont {S.~V.}\ \bibnamefont
  {Eremeev}}, \bibinfo {author} {\bibfnamefont {E.~E.}\ \bibnamefont
  {Krasovskii}}, \bibinfo {author} {\bibfnamefont {P.~M.}\ \bibnamefont
  {Echenique}},\ and\ \bibinfo {author} {\bibfnamefont {E.~V.}\ \bibnamefont
  {Chulkov}},\ }\bibfield  {title} {\bibinfo {title} {Quantum spin {Hall}
  insulators in centrosymmetric thin films composed from topologically trivial
  {BiTeI} trilayers},\ }\href {https://doi.org/10.1038/srep43666} {\bibfield
  {journal} {\bibinfo  {journal} {Scientific Reports}\ }\textbf {\bibinfo
  {volume} {7}},\ \bibinfo {pages} {43666} (\bibinfo {year}
  {2017})}\BibitemShut {NoStop}%
\bibitem [{\citenamefont {Nechaev}\ and\ \citenamefont
  {Krasovskii}(2018)}]{Nechaev_PRB_2018}%
  \BibitemOpen
  \bibfield  {author} {\bibinfo {author} {\bibfnamefont {I.~A.}\ \bibnamefont
  {Nechaev}}\ and\ \bibinfo {author} {\bibfnamefont {E.~E.}\ \bibnamefont
  {Krasovskii}},\ }\bibfield  {title} {\bibinfo {title} {{Relativistic
  splitting of surface states at Si-terminated surfaces of the layered
  intermetallic compounds $R{T}_{2}{\mathrm{Si}}_{2}$ ($R$=rare earth; $T$=Ir,
  Rh)}},\ }\href {https://doi.org/10.1103/PhysRevB.98.245415} {\bibfield
  {journal} {\bibinfo  {journal} {Phys. Rev. B}\ }\textbf {\bibinfo {volume}
  {98}},\ \bibinfo {pages} {245415} (\bibinfo {year} {2018})}\BibitemShut
  {NoStop}%
\bibitem [{\citenamefont {Schulz}\ \emph {et~al.}(2019)\citenamefont {Schulz},
  \citenamefont {Nechaev}, \citenamefont {G\"{u}ttler}, \citenamefont
  {Poelchen}, \citenamefont {Generalov}, \citenamefont {Danzenb\"{a}cher},
  \citenamefont {Chikina}, \citenamefont {Seiro}, \citenamefont {Kliemt},
  \citenamefont {Vyazovskaya}, \citenamefont {Kim}, \citenamefont {Dudin},
  \citenamefont {Chulkov}, \citenamefont {Laubschat}, \citenamefont
  {Krasovskii}, \citenamefont {Geibel}, \citenamefont {Krellner}, \citenamefont
  {Kummer},\ and\ \citenamefont {Vyalikh}}]{Susanne2019}%
  \BibitemOpen
  \bibfield  {author} {\bibinfo {author} {\bibfnamefont {S.}~\bibnamefont
  {Schulz}}, \bibinfo {author} {\bibfnamefont {I.~A.}\ \bibnamefont {Nechaev}},
  \bibinfo {author} {\bibfnamefont {M.}~\bibnamefont {G\"{u}ttler}}, \bibinfo
  {author} {\bibfnamefont {G.}~\bibnamefont {Poelchen}}, \bibinfo {author}
  {\bibfnamefont {A.}~\bibnamefont {Generalov}}, \bibinfo {author}
  {\bibfnamefont {S.}~\bibnamefont {Danzenb\"{a}cher}}, \bibinfo {author}
  {\bibfnamefont {A.}~\bibnamefont {Chikina}}, \bibinfo {author} {\bibfnamefont
  {S.}~\bibnamefont {Seiro}}, \bibinfo {author} {\bibfnamefont
  {K.}~\bibnamefont {Kliemt}}, \bibinfo {author} {\bibfnamefont {A.~Y.}\
  \bibnamefont {Vyazovskaya}}, \bibinfo {author} {\bibfnamefont {T.~K.}\
  \bibnamefont {Kim}}, \bibinfo {author} {\bibfnamefont {P.}~\bibnamefont
  {Dudin}}, \bibinfo {author} {\bibfnamefont {E.~V.}\ \bibnamefont {Chulkov}},
  \bibinfo {author} {\bibfnamefont {C.}~\bibnamefont {Laubschat}}, \bibinfo
  {author} {\bibfnamefont {E.~E.}\ \bibnamefont {Krasovskii}}, \bibinfo
  {author} {\bibfnamefont {C.}~\bibnamefont {Geibel}}, \bibinfo {author}
  {\bibfnamefont {C.}~\bibnamefont {Krellner}}, \bibinfo {author}
  {\bibfnamefont {K.}~\bibnamefont {Kummer}},\ and\ \bibinfo {author}
  {\bibfnamefont {D.~V.}\ \bibnamefont {Vyalikh}},\ }\bibfield  {title}
  {\bibinfo {title} {{Emerging 2D-ferromagnetism and strong spin-orbit coupling
  at the surface of valence-fluctuating EuIr$_2$Si$_2$}},\ }\href
  {https://doi.org/10.1038/s41535-019-0166-z} {\bibfield  {journal} {\bibinfo
  {journal} {npj Quantum Mater.}\ }\textbf {\bibinfo {volume} {4}},\ \bibinfo
  {pages} {26} (\bibinfo {year} {2019})}\BibitemShut {NoStop}%
\bibitem [{\citenamefont {Usachov}\ \emph {et~al.}(2020)\citenamefont
  {Usachov}, \citenamefont {Nechaev}, \citenamefont {Poelchen}, \citenamefont
  {G\"uttler}, \citenamefont {Krasovskii}, \citenamefont {Schulz},
  \citenamefont {Generalov}, \citenamefont {Kliemt}, \citenamefont {Kraiker},
  \citenamefont {Krellner}, \citenamefont {Kummer}, \citenamefont
  {Danzenb\"acher}, \citenamefont {Laubschat}, \citenamefont {Weber},
  \citenamefont {S\'anchez-Barriga}, \citenamefont {Chulkov}, \citenamefont
  {Santander-Syro}, \citenamefont {Imai}, \citenamefont {Miyamoto},
  \citenamefont {Okuda},\ and\ \citenamefont {Vyalikh}}]{Usachov_PRL_2020}%
  \BibitemOpen
  \bibfield  {author} {\bibinfo {author} {\bibfnamefont {D.~Y.}\ \bibnamefont
  {Usachov}}, \bibinfo {author} {\bibfnamefont {I.~A.}\ \bibnamefont
  {Nechaev}}, \bibinfo {author} {\bibfnamefont {G.}~\bibnamefont {Poelchen}},
  \bibinfo {author} {\bibfnamefont {M.}~\bibnamefont {G\"uttler}}, \bibinfo
  {author} {\bibfnamefont {E.~E.}\ \bibnamefont {Krasovskii}}, \bibinfo
  {author} {\bibfnamefont {S.}~\bibnamefont {Schulz}}, \bibinfo {author}
  {\bibfnamefont {A.}~\bibnamefont {Generalov}}, \bibinfo {author}
  {\bibfnamefont {K.}~\bibnamefont {Kliemt}}, \bibinfo {author} {\bibfnamefont
  {A.}~\bibnamefont {Kraiker}}, \bibinfo {author} {\bibfnamefont
  {C.}~\bibnamefont {Krellner}}, \bibinfo {author} {\bibfnamefont
  {K.}~\bibnamefont {Kummer}}, \bibinfo {author} {\bibfnamefont
  {S.}~\bibnamefont {Danzenb\"acher}}, \bibinfo {author} {\bibfnamefont
  {C.}~\bibnamefont {Laubschat}}, \bibinfo {author} {\bibfnamefont {A.~P.}\
  \bibnamefont {Weber}}, \bibinfo {author} {\bibfnamefont {J.}~\bibnamefont
  {S\'anchez-Barriga}}, \bibinfo {author} {\bibfnamefont {E.~V.}\ \bibnamefont
  {Chulkov}}, \bibinfo {author} {\bibfnamefont {A.~F.}\ \bibnamefont
  {Santander-Syro}}, \bibinfo {author} {\bibfnamefont {T.}~\bibnamefont
  {Imai}}, \bibinfo {author} {\bibfnamefont {K.}~\bibnamefont {Miyamoto}},
  \bibinfo {author} {\bibfnamefont {T.}~\bibnamefont {Okuda}},\ and\ \bibinfo
  {author} {\bibfnamefont {D.~V.}\ \bibnamefont {Vyalikh}},\ }\bibfield
  {title} {\bibinfo {title} {{Cubic Rashba Effect in the Surface Spin Structure
  of Rare-Earth Ternary Materials}},\ }\href
  {https://doi.org/10.1103/PhysRevLett.124.237202} {\bibfield  {journal}
  {\bibinfo  {journal} {Phys. Rev. Lett.}\ }\textbf {\bibinfo {volume} {124}},\
  \bibinfo {pages} {237202} (\bibinfo {year} {2020})}\BibitemShut {NoStop}%
\bibitem [{\citenamefont {Nechaev}\ and\ \citenamefont
  {Krasovskii}(2019)}]{Nechaev_PRB_2019}%
  \BibitemOpen
  \bibfield  {author} {\bibinfo {author} {\bibfnamefont {I.~A.}\ \bibnamefont
  {Nechaev}}\ and\ \bibinfo {author} {\bibfnamefont {E.~E.}\ \bibnamefont
  {Krasovskii}},\ }\bibfield  {title} {\bibinfo {title} {{Spin polarization by
  first-principles relativistic $\mathrm{k}\cdot\mathrm{p}$ theory: Application
  to the surface alloys ${\mathrm{PbAg}}_{2}$ and ${\mathrm{BiAg}}_{2}$}},\
  }\href {https://doi.org/10.1103/PhysRevB.100.115432} {\bibfield  {journal}
  {\bibinfo  {journal} {Phys. Rev. B}\ }\textbf {\bibinfo {volume} {100}},\
  \bibinfo {pages} {115432} (\bibinfo {year} {2019})}\BibitemShut {NoStop}%
\bibitem [{\citenamefont {Nechaev}\ and\ \citenamefont
  {Krasovskii}(2020)}]{Nechaev_PRB_2020}%
  \BibitemOpen
  \bibfield  {author} {\bibinfo {author} {\bibfnamefont {I.~A.}\ \bibnamefont
  {Nechaev}}\ and\ \bibinfo {author} {\bibfnamefont {E.~E.}\ \bibnamefont
  {Krasovskii}},\ }\bibfield  {title} {\bibinfo {title} {{Ab initio k
  $\ifmmode\cdot\else\textperiodcentered\fi{}$ p theory of spin-momentum
  locking: Application to topological surface states}},\ }\href
  {https://doi.org/10.1103/PhysRevB.102.115437} {\bibfield  {journal} {\bibinfo
   {journal} {Phys. Rev. B}\ }\textbf {\bibinfo {volume} {102}},\ \bibinfo
  {pages} {115437} (\bibinfo {year} {2020})}\BibitemShut {NoStop}%
\bibitem [{\citenamefont {Liu}\ \emph {et~al.}(2010)\citenamefont {Liu},
  \citenamefont {Qi}, \citenamefont {Zhang}, \citenamefont {Dai}, \citenamefont
  {Fang},\ and\ \citenamefont {Zhang}}]{Liu_PRB_2010}%
  \BibitemOpen
  \bibfield  {author} {\bibinfo {author} {\bibfnamefont {C.-X.}\ \bibnamefont
  {Liu}}, \bibinfo {author} {\bibfnamefont {X.-L.}\ \bibnamefont {Qi}},
  \bibinfo {author} {\bibfnamefont {H.}~\bibnamefont {Zhang}}, \bibinfo
  {author} {\bibfnamefont {X.}~\bibnamefont {Dai}}, \bibinfo {author}
  {\bibfnamefont {Z.}~\bibnamefont {Fang}},\ and\ \bibinfo {author}
  {\bibfnamefont {S.-C.}\ \bibnamefont {Zhang}},\ }\bibfield  {title} {\bibinfo
  {title} {{Model Hamiltonian for topological insulators}},\ }\href
  {https://doi.org/10.1103/PhysRevB.82.045122} {\bibfield  {journal} {\bibinfo
  {journal} {Phys. Rev. B}\ }\textbf {\bibinfo {volume} {82}},\ \bibinfo
  {pages} {045122} (\bibinfo {year} {2010})}\BibitemShut {NoStop}%
\bibitem [{\citenamefont {Krasovskii}(1997)}]{Krasovskii_PRB_1997}%
  \BibitemOpen
  \bibfield  {author} {\bibinfo {author} {\bibfnamefont {E.~E.}\ \bibnamefont
  {Krasovskii}},\ }\bibfield  {title} {\bibinfo {title} {Accuracy and
  convergence properties of the extended linear augmented-plane-wave method},\
  }\href {https://doi.org/10.1103/PhysRevB.56.12866} {\bibfield  {journal}
  {\bibinfo  {journal} {Phys. Rev. B}\ }\textbf {\bibinfo {volume} {56}},\
  \bibinfo {pages} {12866} (\bibinfo {year} {1997})}\BibitemShut {NoStop}%
\bibitem [{\citenamefont {Krasovskii}\ \emph {et~al.}(1999)\citenamefont
  {Krasovskii}, \citenamefont {Starrost},\ and\ \citenamefont
  {Schattke}}]{Krasovskii_PRB_1999}%
  \BibitemOpen
  \bibfield  {author} {\bibinfo {author} {\bibfnamefont {E.~E.}\ \bibnamefont
  {Krasovskii}}, \bibinfo {author} {\bibfnamefont {F.}~\bibnamefont
  {Starrost}},\ and\ \bibinfo {author} {\bibfnamefont {W.}~\bibnamefont
  {Schattke}},\ }\bibfield  {title} {\bibinfo {title} {Augmented fourier
  components method for constructing the crystal potential in self-consistent
  band-structure calculations},\ }\href
  {https://doi.org/10.1103/PhysRevB.59.10504} {\bibfield  {journal} {\bibinfo
  {journal} {Phys. Rev. B}\ }\textbf {\bibinfo {volume} {59}},\ \bibinfo
  {pages} {10504} (\bibinfo {year} {1999})}\BibitemShut {NoStop}%
\bibitem [{\citenamefont {Koelling}\ and\ \citenamefont
  {Harmon}(1977)}]{Koelling_1977}%
  \BibitemOpen
  \bibfield  {author} {\bibinfo {author} {\bibfnamefont {D.~D.}\ \bibnamefont
  {Koelling}}\ and\ \bibinfo {author} {\bibfnamefont {B.~N.}\ \bibnamefont
  {Harmon}},\ }\bibfield  {title} {\bibinfo {title} {A technique for
  relativistic spin-polarised calculations},\ }\href
  {https://doi.org/10.1088/0022-3719/10/16/019} {\bibfield  {journal} {\bibinfo
   {journal} {Journal of Physics C: Solid State Physics}\ }\textbf {\bibinfo
  {volume} {10}},\ \bibinfo {pages} {3107} (\bibinfo {year}
  {1977})}\BibitemShut {NoStop}%
\bibitem [{\citenamefont {Wyckoff}(1964)}]{Wyckoff_RWG}%
  \BibitemOpen
  \bibfield  {author} {\bibinfo {author} {\bibfnamefont {R.~W.~G.}\
  \bibnamefont {Wyckoff}},\ }\href@noop {} {\emph {\bibinfo {title} {{Crystal
  Structures 2}}}}\ (\bibinfo  {publisher} {John Wiley and Sons},\ \bibinfo
  {address} {New York},\ \bibinfo {year} {1964})\BibitemShut {NoStop}%
\bibitem [{\citenamefont {Nechaev}\ \emph {et~al.}(2013)\citenamefont
  {Nechaev}, \citenamefont {Hatch}, \citenamefont {Bianchi}, \citenamefont
  {Guan}, \citenamefont {Friedrich}, \citenamefont {Aguilera}, \citenamefont
  {Mi}, \citenamefont {Iversen}, \citenamefont {Bl\"ugel}, \citenamefont
  {Hofmann},\ and\ \citenamefont {Chulkov}}]{Nechaev_PRB_2013_BISE}%
  \BibitemOpen
  \bibfield  {author} {\bibinfo {author} {\bibfnamefont {I.~A.}\ \bibnamefont
  {Nechaev}}, \bibinfo {author} {\bibfnamefont {R.~C.}\ \bibnamefont {Hatch}},
  \bibinfo {author} {\bibfnamefont {M.}~\bibnamefont {Bianchi}}, \bibinfo
  {author} {\bibfnamefont {D.}~\bibnamefont {Guan}}, \bibinfo {author}
  {\bibfnamefont {C.}~\bibnamefont {Friedrich}}, \bibinfo {author}
  {\bibfnamefont {I.}~\bibnamefont {Aguilera}}, \bibinfo {author}
  {\bibfnamefont {J.~L.}\ \bibnamefont {Mi}}, \bibinfo {author} {\bibfnamefont
  {B.~B.}\ \bibnamefont {Iversen}}, \bibinfo {author} {\bibfnamefont
  {S.}~\bibnamefont {Bl\"ugel}}, \bibinfo {author} {\bibfnamefont
  {P.}~\bibnamefont {Hofmann}},\ and\ \bibinfo {author} {\bibfnamefont {E.~V.}\
  \bibnamefont {Chulkov}},\ }\bibfield  {title} {\bibinfo {title} {{Evidence
  for a direct band gap in the topological insulator Bi${}_{2}$Se${}_{3}$ from
  theory and experiment}},\ }\href {https://doi.org/10.1103/PhysRevB.87.121111}
  {\bibfield  {journal} {\bibinfo  {journal} {Phys. Rev. B}\ }\textbf {\bibinfo
  {volume} {87}},\ \bibinfo {pages} {121111} (\bibinfo {year}
  {2013})}\BibitemShut {NoStop}%
\bibitem [{\citenamefont {Nechaev}\ and\ \citenamefont
  {Chulkov}(2013)}]{Nechaev_PRB_2013_BITE}%
  \BibitemOpen
  \bibfield  {author} {\bibinfo {author} {\bibfnamefont {I.~A.}\ \bibnamefont
  {Nechaev}}\ and\ \bibinfo {author} {\bibfnamefont {E.~V.}\ \bibnamefont
  {Chulkov}},\ }\bibfield  {title} {\bibinfo {title} {{Quasiparticle band gap
  in the topological insulator Bi${}_{2}$Te${}_{3}$}},\ }\href
  {https://doi.org/10.1103/PhysRevB.88.165135} {\bibfield  {journal} {\bibinfo
  {journal} {Phys. Rev. B}\ }\textbf {\bibinfo {volume} {88}},\ \bibinfo
  {pages} {165135} (\bibinfo {year} {2013})}\BibitemShut {NoStop}%
\bibitem [{\citenamefont {Nechaev}\ \emph {et~al.}(2015)\citenamefont
  {Nechaev}, \citenamefont {Aguilera}, \citenamefont {De~Renzi}, \citenamefont
  {di~Bona}, \citenamefont {Lodi~Rizzini}, \citenamefont {Mio}, \citenamefont
  {Nicotra}, \citenamefont {Politano}, \citenamefont {Scalese}, \citenamefont
  {Aliev}, \citenamefont {Babanly}, \citenamefont {Friedrich}, \citenamefont
  {Bl\"ugel},\ and\ \citenamefont {Chulkov}}]{Nechaev_PRB_2015_SBTE}%
  \BibitemOpen
  \bibfield  {author} {\bibinfo {author} {\bibfnamefont {I.~A.}\ \bibnamefont
  {Nechaev}}, \bibinfo {author} {\bibfnamefont {I.}~\bibnamefont {Aguilera}},
  \bibinfo {author} {\bibfnamefont {V.}~\bibnamefont {De~Renzi}}, \bibinfo
  {author} {\bibfnamefont {A.}~\bibnamefont {di~Bona}}, \bibinfo {author}
  {\bibfnamefont {A.}~\bibnamefont {Lodi~Rizzini}}, \bibinfo {author}
  {\bibfnamefont {A.~M.}\ \bibnamefont {Mio}}, \bibinfo {author} {\bibfnamefont
  {G.}~\bibnamefont {Nicotra}}, \bibinfo {author} {\bibfnamefont
  {A.}~\bibnamefont {Politano}}, \bibinfo {author} {\bibfnamefont
  {S.}~\bibnamefont {Scalese}}, \bibinfo {author} {\bibfnamefont {Z.~S.}\
  \bibnamefont {Aliev}}, \bibinfo {author} {\bibfnamefont {M.~B.}\ \bibnamefont
  {Babanly}}, \bibinfo {author} {\bibfnamefont {C.}~\bibnamefont {Friedrich}},
  \bibinfo {author} {\bibfnamefont {S.}~\bibnamefont {Bl\"ugel}},\ and\
  \bibinfo {author} {\bibfnamefont {E.~V.}\ \bibnamefont {Chulkov}},\
  }\bibfield  {title} {\bibinfo {title} {{Quasiparticle spectrum and plasmonic
  excitations in the topological insulator
  ${\mathrm{Sb}}_{2}{\mathrm{Te}}_{3}$}},\ }\href
  {https://doi.org/10.1103/PhysRevB.91.245123} {\bibfield  {journal} {\bibinfo
  {journal} {Phys. Rev. B}\ }\textbf {\bibinfo {volume} {91}},\ \bibinfo
  {pages} {245123} (\bibinfo {year} {2015})}\BibitemShut {NoStop}%
\bibitem [{\citenamefont {L\"{o}wdin}(1951)}]{Leowdin_JCP_1951}%
  \BibitemOpen
  \bibfield  {author} {\bibinfo {author} {\bibfnamefont {P.-O.}\ \bibnamefont
  {L\"{o}wdin}},\ }\bibfield  {title} {\bibinfo {title} {{A Note on the
  Quantum-Mechanical Perturbation Theory}},\ }\href
  {https://doi.org/10.1063/1.1748067} {\bibfield  {journal} {\bibinfo
  {journal} {The Journal of Chemical Physics}\ }\textbf {\bibinfo {volume}
  {19}},\ \bibinfo {pages} {1396} (\bibinfo {year} {1951})}\BibitemShut
  {NoStop}%
\bibitem [{\citenamefont {Schrieffer}\ and\ \citenamefont
  {Wolff}(1966)}]{Schrieffer_PR_1966}%
  \BibitemOpen
  \bibfield  {author} {\bibinfo {author} {\bibfnamefont {J.~R.}\ \bibnamefont
  {Schrieffer}}\ and\ \bibinfo {author} {\bibfnamefont {P.~A.}\ \bibnamefont
  {Wolff}},\ }\bibfield  {title} {\bibinfo {title} {{Relation between the
  Anderson and Kondo Hamiltonians}},\ }\href
  {https://doi.org/10.1103/PhysRev.149.491} {\bibfield  {journal} {\bibinfo
  {journal} {Phys. Rev.}\ }\textbf {\bibinfo {volume} {149}},\ \bibinfo {pages}
  {491} (\bibinfo {year} {1966})}\BibitemShut {NoStop}%
\bibitem [{\citenamefont {Winkler}(2003)}]{Winkler_KP}%
  \BibitemOpen
  \bibfield  {author} {\bibinfo {author} {\bibfnamefont {R.}~\bibnamefont
  {Winkler}},\ }\href@noop {} {\emph {\bibinfo {title} {{Spin-Orbit Coupling
  Effects in Two-Dimensional Electron and Hole Systems}}}}\ (\bibinfo
  {publisher} {Springer},\ \bibinfo {address} {Berlin},\ \bibinfo {year}
  {2003})\BibitemShut {NoStop}%
\bibitem [{\citenamefont {Silvestrov}\ \emph {et~al.}(2012)\citenamefont
  {Silvestrov}, \citenamefont {Brouwer},\ and\ \citenamefont
  {Mishchenko}}]{Silvestrov_PRB_2012}%
  \BibitemOpen
  \bibfield  {author} {\bibinfo {author} {\bibfnamefont {P.~G.}\ \bibnamefont
  {Silvestrov}}, \bibinfo {author} {\bibfnamefont {P.~W.}\ \bibnamefont
  {Brouwer}},\ and\ \bibinfo {author} {\bibfnamefont {E.~G.}\ \bibnamefont
  {Mishchenko}},\ }\bibfield  {title} {\bibinfo {title} {Spin and charge
  structure of the surface states in topological insulators},\ }\href
  {https://doi.org/10.1103/PhysRevB.86.075302} {\bibfield  {journal} {\bibinfo
  {journal} {Phys. Rev. B}\ }\textbf {\bibinfo {volume} {86}},\ \bibinfo
  {pages} {075302} (\bibinfo {year} {2012})}\BibitemShut {NoStop}%
\bibitem [{\citenamefont {Brey}\ and\ \citenamefont
  {Fertig}(2014)}]{Brey_PRB_2014}%
  \BibitemOpen
  \bibfield  {author} {\bibinfo {author} {\bibfnamefont {L.}~\bibnamefont
  {Brey}}\ and\ \bibinfo {author} {\bibfnamefont {H.~A.}\ \bibnamefont
  {Fertig}},\ }\bibfield  {title} {\bibinfo {title} {{Electronic states of
  wires and slabs of topological insulators: Quantum Hall effects and edge
  transport}},\ }\href {https://doi.org/10.1103/PhysRevB.89.085305} {\bibfield
  {journal} {\bibinfo  {journal} {Phys. Rev. B}\ }\textbf {\bibinfo {volume}
  {89}},\ \bibinfo {pages} {085305} (\bibinfo {year} {2014})}\BibitemShut
  {NoStop}%
\bibitem [{\citenamefont {Eppenga}\ \emph {et~al.}(1987)\citenamefont
  {Eppenga}, \citenamefont {Schuurmans},\ and\ \citenamefont
  {Colak}}]{Eppenga_PRB_1987}%
  \BibitemOpen
  \bibfield  {author} {\bibinfo {author} {\bibfnamefont {R.}~\bibnamefont
  {Eppenga}}, \bibinfo {author} {\bibfnamefont {M.~F.~H.}\ \bibnamefont
  {Schuurmans}},\ and\ \bibinfo {author} {\bibfnamefont {S.}~\bibnamefont
  {Colak}},\ }\bibfield  {title} {\bibinfo {title} {{New k\ensuremath{\cdot}p
  theory for GaAs/${\mathrm{Ga}}_{1\mathrm{\ensuremath{-}}\mathrm{x}}$
  ${\mathrm{Al}}_{\mathrm{x}}$As-type quantum wells}},\ }\href
  {https://doi.org/10.1103/PhysRevB.36.1554} {\bibfield  {journal} {\bibinfo
  {journal} {Phys. Rev. B}\ }\textbf {\bibinfo {volume} {36}},\ \bibinfo
  {pages} {1554} (\bibinfo {year} {1987})}\BibitemShut {NoStop}%
\bibitem [{\citenamefont {Foreman}(1993)}]{Foreman_PRB_1993}%
  \BibitemOpen
  \bibfield  {author} {\bibinfo {author} {\bibfnamefont {B.~A.}\ \bibnamefont
  {Foreman}},\ }\bibfield  {title} {\bibinfo {title} {{Effective-mass
  Hamiltonian and boundary conditions for the valence bands of semiconductor
  microstructures}},\ }\href {https://doi.org/10.1103/PhysRevB.48.4964}
  {\bibfield  {journal} {\bibinfo  {journal} {Phys. Rev. B}\ }\textbf {\bibinfo
  {volume} {48}},\ \bibinfo {pages} {4964} (\bibinfo {year}
  {1993})}\BibitemShut {NoStop}%
\bibitem [{\citenamefont {Dargys}(2007)}]{Dargys_2007}%
  \BibitemOpen
  \bibfield  {author} {\bibinfo {author} {\bibfnamefont {A.}~\bibnamefont
  {Dargys}},\ }\bibfield  {title} {\bibinfo {title} {Spin and orbital motion
  surfaces in {HgTe}},\ }\href {https://doi.org/10.1088/0268-1242/22/5/007}
  {\bibfield  {journal} {\bibinfo  {journal} {Semiconductor Science and
  Technology}\ }\textbf {\bibinfo {volume} {22}},\ \bibinfo {pages} {497}
  (\bibinfo {year} {2007})}\BibitemShut {NoStop}%
\bibitem [{\citenamefont {Abolfath}\ \emph {et~al.}(2001)\citenamefont
  {Abolfath}, \citenamefont {Jungwirth}, \citenamefont {Brum},\ and\
  \citenamefont {MacDonald}}]{Abolfath_PRB_2011}%
  \BibitemOpen
  \bibfield  {author} {\bibinfo {author} {\bibfnamefont {M.}~\bibnamefont
  {Abolfath}}, \bibinfo {author} {\bibfnamefont {T.}~\bibnamefont {Jungwirth}},
  \bibinfo {author} {\bibfnamefont {J.}~\bibnamefont {Brum}},\ and\ \bibinfo
  {author} {\bibfnamefont {A.~H.}\ \bibnamefont {MacDonald}},\ }\bibfield
  {title} {\bibinfo {title} {{Theory of magnetic anisotropy in
  ${\mathrm{III}}_{1\ensuremath{-}x}{\mathrm{Mn}}_{x}\mathrm{V}$
  ferromagnets}},\ }\href {https://doi.org/10.1103/PhysRevB.63.054418}
  {\bibfield  {journal} {\bibinfo  {journal} {Phys. Rev. B}\ }\textbf {\bibinfo
  {volume} {63}},\ \bibinfo {pages} {054418} (\bibinfo {year}
  {2001})}\BibitemShut {NoStop}%
\bibitem [{\citenamefont {Wang}\ \emph {et~al.}(2006)\citenamefont {Wang},
  \citenamefont {Yates}, \citenamefont {Souza},\ and\ \citenamefont
  {Vanderbilt}}]{Wang_PRB_2006}%
  \BibitemOpen
  \bibfield  {author} {\bibinfo {author} {\bibfnamefont {X.}~\bibnamefont
  {Wang}}, \bibinfo {author} {\bibfnamefont {J.~R.}\ \bibnamefont {Yates}},
  \bibinfo {author} {\bibfnamefont {I.}~\bibnamefont {Souza}},\ and\ \bibinfo
  {author} {\bibfnamefont {D.}~\bibnamefont {Vanderbilt}},\ }\bibfield  {title}
  {\bibinfo {title} {{Ab initio calculation of the anomalous Hall conductivity
  by Wannier interpolation}},\ }\href
  {https://doi.org/10.1103/PhysRevB.74.195118} {\bibfield  {journal} {\bibinfo
  {journal} {Phys. Rev. B}\ }\textbf {\bibinfo {volume} {74}},\ \bibinfo
  {pages} {195118} (\bibinfo {year} {2006})}\BibitemShut {NoStop}%
\bibitem [{\citenamefont {Fang}\ \emph {et~al.}(2014)\citenamefont {Fang},
  \citenamefont {Gilbert},\ and\ \citenamefont {Bernevig}}]{Fang_PRL_2014}%
  \BibitemOpen
  \bibfield  {author} {\bibinfo {author} {\bibfnamefont {C.}~\bibnamefont
  {Fang}}, \bibinfo {author} {\bibfnamefont {M.~J.}\ \bibnamefont {Gilbert}},\
  and\ \bibinfo {author} {\bibfnamefont {B.~A.}\ \bibnamefont {Bernevig}},\
  }\bibfield  {title} {\bibinfo {title} {{Large-Chern-Number Quantum Anomalous
  Hall Effect in Thin-Film Topological Crystalline Insulators}},\ }\href
  {https://doi.org/10.1103/PhysRevLett.112.046801} {\bibfield  {journal}
  {\bibinfo  {journal} {Phys. Rev. Lett.}\ }\textbf {\bibinfo {volume} {112}},\
  \bibinfo {pages} {046801} (\bibinfo {year} {2014})}\BibitemShut {NoStop}%
\bibitem [{\citenamefont {Onoda}\ and\ \citenamefont
  {Nagaosa}(2002)}]{Onoda_JPSJ_2002}%
  \BibitemOpen
  \bibfield  {author} {\bibinfo {author} {\bibfnamefont {M.}~\bibnamefont
  {Onoda}}\ and\ \bibinfo {author} {\bibfnamefont {N.}~\bibnamefont
  {Nagaosa}},\ }\bibfield  {title} {\bibinfo {title} {{Topological Nature of
  Anomalous Hall Effect in Ferromagnets}},\ }\href
  {https://doi.org/10.1143/JPSJ.71.19} {\bibfield  {journal} {\bibinfo
  {journal} {Journal of the Physical Society of Japan}\ }\textbf {\bibinfo
  {volume} {71}},\ \bibinfo {pages} {19} (\bibinfo {year} {2002})}\BibitemShut
  {NoStop}%
\bibitem [{\citenamefont {Xu}\ \emph {et~al.}(2011)\citenamefont {Xu},
  \citenamefont {Weng}, \citenamefont {Wang}, \citenamefont {Dai},\ and\
  \citenamefont {Fang}}]{Xu_PRL_2011}%
  \BibitemOpen
  \bibfield  {author} {\bibinfo {author} {\bibfnamefont {G.}~\bibnamefont
  {Xu}}, \bibinfo {author} {\bibfnamefont {H.}~\bibnamefont {Weng}}, \bibinfo
  {author} {\bibfnamefont {Z.}~\bibnamefont {Wang}}, \bibinfo {author}
  {\bibfnamefont {X.}~\bibnamefont {Dai}},\ and\ \bibinfo {author}
  {\bibfnamefont {Z.}~\bibnamefont {Fang}},\ }\bibfield  {title} {\bibinfo
  {title} {{Chern Semimetal and the Quantized Anomalous Hall Effect in
  ${\mathrm{HgCr}}_{2}{\mathrm{Se}}_{4}$}},\ }\href
  {https://doi.org/10.1103/PhysRevLett.107.186806} {\bibfield  {journal}
  {\bibinfo  {journal} {Phys. Rev. Lett.}\ }\textbf {\bibinfo {volume} {107}},\
  \bibinfo {pages} {186806} (\bibinfo {year} {2011})}\BibitemShut {NoStop}%
\bibitem [{\citenamefont {Liu}\ \emph {et~al.}(2018)\citenamefont {Liu},
  \citenamefont {Sun}, \citenamefont {Kumar}, \citenamefont {Muechler},
  \citenamefont {Sun}, \citenamefont {Jiao}, \citenamefont {Yang},
  \citenamefont {Liu}, \citenamefont {Liang}, \citenamefont {Xu}, \citenamefont
  {Kroder}, \citenamefont {S\"{u}{\ss}}, \citenamefont {Borrmann},
  \citenamefont {Shekhar}, \citenamefont {Wang}, \citenamefont {Xi},
  \citenamefont {Wang}, \citenamefont {Schnelle}, \citenamefont {Wirth},
  \citenamefont {Chen}, \citenamefont {Goennenwein},\ and\ \citenamefont
  {Felser}}]{Liu_NatPhys_2018}%
  \BibitemOpen
  \bibfield  {author} {\bibinfo {author} {\bibfnamefont {E.}~\bibnamefont
  {Liu}}, \bibinfo {author} {\bibfnamefont {Y.}~\bibnamefont {Sun}}, \bibinfo
  {author} {\bibfnamefont {N.}~\bibnamefont {Kumar}}, \bibinfo {author}
  {\bibfnamefont {L.}~\bibnamefont {Muechler}}, \bibinfo {author}
  {\bibfnamefont {A.}~\bibnamefont {Sun}}, \bibinfo {author} {\bibfnamefont
  {L.}~\bibnamefont {Jiao}}, \bibinfo {author} {\bibfnamefont {S.-Y.}\
  \bibnamefont {Yang}}, \bibinfo {author} {\bibfnamefont {D.}~\bibnamefont
  {Liu}}, \bibinfo {author} {\bibfnamefont {A.}~\bibnamefont {Liang}}, \bibinfo
  {author} {\bibfnamefont {Q.}~\bibnamefont {Xu}}, \bibinfo {author}
  {\bibfnamefont {J.}~\bibnamefont {Kroder}}, \bibinfo {author} {\bibfnamefont
  {V.}~\bibnamefont {S\"{u}{\ss}}}, \bibinfo {author} {\bibfnamefont
  {H.}~\bibnamefont {Borrmann}}, \bibinfo {author} {\bibfnamefont
  {C.}~\bibnamefont {Shekhar}}, \bibinfo {author} {\bibfnamefont
  {Z.}~\bibnamefont {Wang}}, \bibinfo {author} {\bibfnamefont {C.}~\bibnamefont
  {Xi}}, \bibinfo {author} {\bibfnamefont {W.}~\bibnamefont {Wang}}, \bibinfo
  {author} {\bibfnamefont {W.}~\bibnamefont {Schnelle}}, \bibinfo {author}
  {\bibfnamefont {S.}~\bibnamefont {Wirth}}, \bibinfo {author} {\bibfnamefont
  {Y.}~\bibnamefont {Chen}}, \bibinfo {author} {\bibfnamefont {S.~T.~B.}\
  \bibnamefont {Goennenwein}},\ and\ \bibinfo {author} {\bibfnamefont
  {C.}~\bibnamefont {Felser}},\ }\bibfield  {title} {\bibinfo {title} {{Giant
  anomalous Hall effect in a ferromagnetic kagome-lattice semimetal}},\ }\href
  {https://doi.org/10.1038/s41567-018-0234-5} {\bibfield  {journal} {\bibinfo
  {journal} {Nature Physics}\ }\textbf {\bibinfo {volume} {14}},\ \bibinfo
  {pages} {1125} (\bibinfo {year} {2018})}\BibitemShut {NoStop}%
\bibitem [{\citenamefont {Jin}\ \emph {et~al.}(2011)\citenamefont {Jin},
  \citenamefont {Im},\ and\ \citenamefont {Freeman}}]{Jin_PRB_2011}%
  \BibitemOpen
  \bibfield  {author} {\bibinfo {author} {\bibfnamefont {H.}~\bibnamefont
  {Jin}}, \bibinfo {author} {\bibfnamefont {J.}~\bibnamefont {Im}},\ and\
  \bibinfo {author} {\bibfnamefont {A.~J.}\ \bibnamefont {Freeman}},\
  }\bibfield  {title} {\bibinfo {title} {{Topological and magnetic phase
  transitions in Bi${}_{2}$Se${}_{3}$ thin films with magnetic impurities}},\
  }\href {https://doi.org/10.1103/PhysRevB.84.134408} {\bibfield  {journal}
  {\bibinfo  {journal} {Phys. Rev. B}\ }\textbf {\bibinfo {volume} {84}},\
  \bibinfo {pages} {134408} (\bibinfo {year} {2011})}\BibitemShut {NoStop}%
\bibitem [{\citenamefont {Muechler}\ \emph {et~al.}(2020)\citenamefont
  {Muechler}, \citenamefont {Liu}, \citenamefont {Gayles}, \citenamefont {Xu},
  \citenamefont {Felser},\ and\ \citenamefont {Sun}}]{Muechler_PRB_2020}%
  \BibitemOpen
  \bibfield  {author} {\bibinfo {author} {\bibfnamefont {L.}~\bibnamefont
  {Muechler}}, \bibinfo {author} {\bibfnamefont {E.}~\bibnamefont {Liu}},
  \bibinfo {author} {\bibfnamefont {J.}~\bibnamefont {Gayles}}, \bibinfo
  {author} {\bibfnamefont {Q.}~\bibnamefont {Xu}}, \bibinfo {author}
  {\bibfnamefont {C.}~\bibnamefont {Felser}},\ and\ \bibinfo {author}
  {\bibfnamefont {Y.}~\bibnamefont {Sun}},\ }\bibfield  {title} {\bibinfo
  {title} {{Emerging chiral edge states from the confinement of a magnetic Weyl
  semimetal in ${\mathrm{Co}}_{3}{\mathrm{Sn}}_{2}{\mathrm{S}}_{2}$}},\ }\href
  {https://doi.org/10.1103/PhysRevB.101.115106} {\bibfield  {journal} {\bibinfo
   {journal} {Phys. Rev. B}\ }\textbf {\bibinfo {volume} {101}},\ \bibinfo
  {pages} {115106} (\bibinfo {year} {2020})}\BibitemShut {NoStop}%
\end{thebibliography}
\end{document}